\documentclass[sigconf,screen,authorversion,nonacm]{acmart}
\setcopyright{none}

\renewcommand\footnotetextcopyrightpermission[1]{}
\AtBeginDocument{%
  }

\setcopyright{none}
\acmISBN{978-1-4503-XXXX-X/2018/06}

\usepackage{fvextra}
\RecustomVerbatimEnvironment{verbatim}{Verbatim}{fontsize=\fontsize{6.65}{7.35}\selectfont,breaklines=true,breakanywhere=true}

\usepackage{float}

\begin{document}

\title{Mixed-Initiative Context: Structuring and Managing Context for Human-AI Collaboration}

\author{Haichang Li}
\affiliation{%
  \institution{George Mason University}
  \city{Fairfax}
  \state{Virginia}
  \country{USA}}

\author{Qinshi Zhang}
\affiliation{%
  \institution{University of California, San Diego}
  \city{San Diego}
  \state{California}
  \country{USA}}

\author{Piaohong Wang}
\affiliation{%
  \institution{City University of Hong Kong}
  \city{Hong Kong}
  \country{China}}

\author{Zhicong Lu}
\affiliation{%
  \institution{George Mason University}
  \city{Fairfax}
  \state{Virginia}
  \country{USA}}

\renewcommand{\shortauthors}{Li et al.}

\begin{abstract}
In the human-AI collaboration area, the context formed naturally through multi-turn interactions is typically flattened into a chronological sequence and treated as a fixed whole in subsequent reasoning, with no mechanism for dynamic organization and management along the collaboration workflow. Yet these contexts differ substantially in lifecycle, structural hierarchy, and relevance. For instance, temporary or abandoned exchanges and parallel topic threads persist in the limited context window, causing interference and even conflict. Meanwhile, users are largely limited to influencing context indirectly through input modifications (e.g., corrections, references, or ignoring), leaving their control neither explicit nor verifiable.

To address this, we propose Mixed-Initiative Context, which reconceptualizes the context formed across multi-turn interactions as an explicit, structured, and manipulable interactive object. Under this concept, the structure, scope, and content of context can be dynamically organized and adjusted according to task needs, enabling both humans and AI to actively participate in context construction and regulation. To explore this concept, we implement Contextify as a probe system and conduct a user study examining users' context management behaviors, attitudes toward AI initiative, and overall collaboration experience. We conclude by discussing the implications of this concept for the HCI community.
\end{abstract}


\begin{teaserfigure}
  \includegraphics[width=\textwidth]{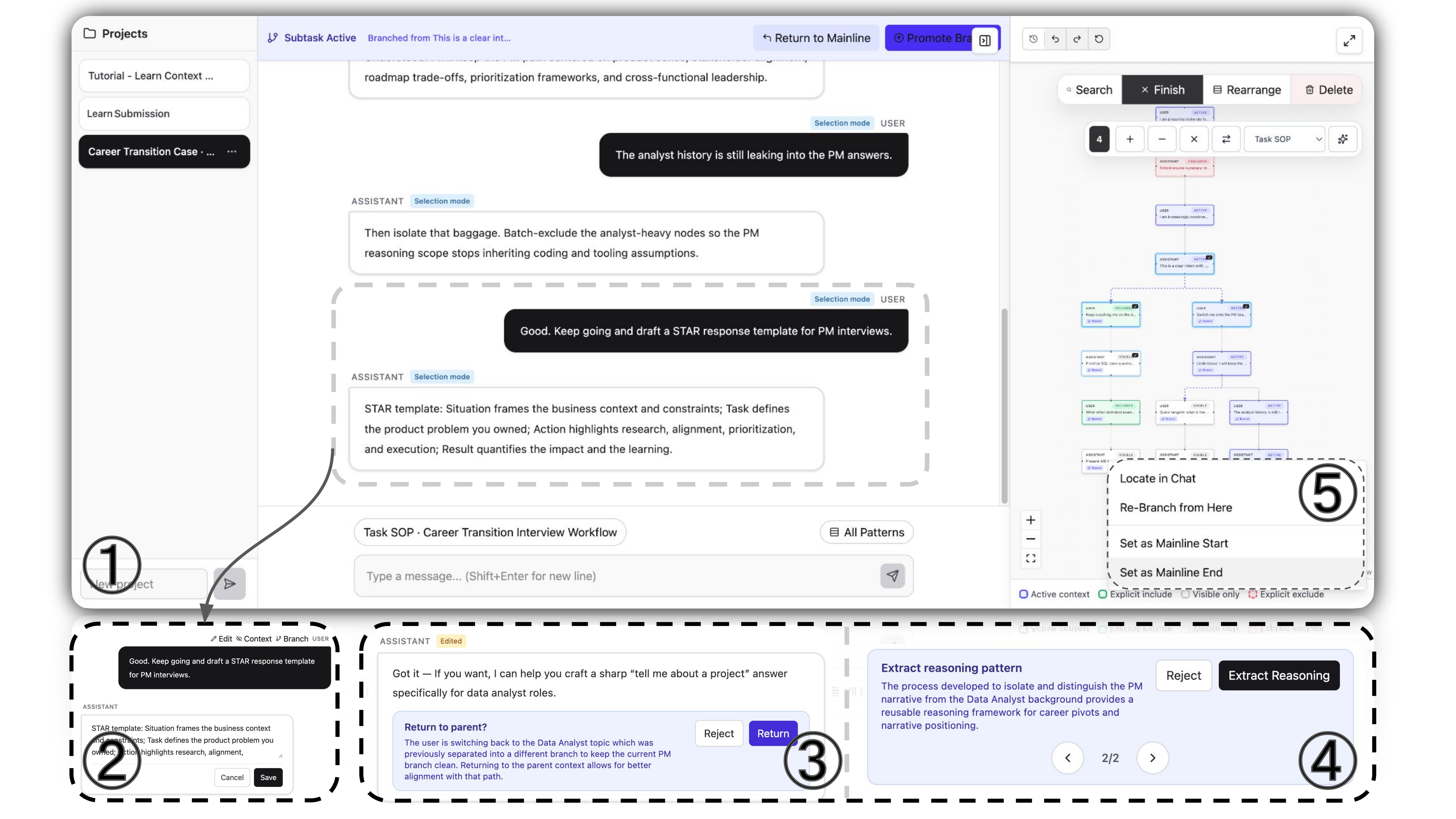}
  \Description{A software interface of the Contextify system featuring three main sections. The left section is a sidebar. The central section is a chat interface showing a conversation flow, with an input box at the bottom topped with pattern capsules (labeled 1). Chat messages display hover menus for editing (labeled 2). Inline cards within the chat suggest returning to a previous branch or extracting a pattern (labeled 3 and 4). The right section displays a vertical tree diagram representing the conversation structure, with a right-click menu open showing options to manipulate the nodes (labeled 5).}
  \caption{Contextify instantiates the Mixed-Initiative Context concept. (1) Conversational System: Top controls navigate or promote mainlines, while input capsules apply patterns. (2) Messages serve as Minimum Context Units supporting direct hover and edition. (3 \& 4) Background and proactive agents analyze interaction traces and user models to proactively recommend structural navigation (3) and pattern extraction (4). (5) Context Map: Synchronized with (1) and (2), it visualizes context topology, allowing users to explicitly include or exclude nodes, set boundaries, and manipulate the hierarchy via right-click menus.}
  \label{fig:teaser} 
\end{teaserfigure}

\maketitle

\section{Introduction}

The long-standing vision of HCI is to build computing systems that understand users' goals, tasks, and current state well enough to provide support at the right moment~\cite{10.1109/MIS.2004.74,10.1145/302979.303030}. For today's large language model (LLM) systems, such support depends on context accumulated across multi-turn interaction: users express goals, add constraints, and revise preferences through ongoing interaction, while AI interprets the current task and infers user intent from that evolving context. Context is therefore not merely a backend implementation detail, but a key substrate for grounding, shared understanding, and continuity in human--AI collaboration~\cite{clark1991grounding,wang2024mutualtheorymindhumanai,liao2023aitransparencyagellms}. As Greenberg and Dourish argue, context is not a static information set, but a dynamic substrate that must be continuously constructed, filtered, and reorganized as tasks progress, collaboration evolves, and user intent shifts~\cite{10.1207/S15327051HCI16234_09,10.1007/s00779-003-0253-8}.

Yet current human--AI systems often treat context as an automatically accumulated, linearly stacked history that is passed wholesale into subsequent reasoning. Users cannot directly inspect what remains active, exclude what is no longer needed, isolate local explorations, or reorganize relationships among pieces of information; instead, they issue new prompts and hope the system infers their updated intent~\cite{10.1145/3746059.3747746,10.1145/3613904.3642462,10.1145/3654777.3676374}. This limitation is especially consequential because user intent is inherently dynamic and human--AI collaboration is rarely linear: design, exploration, and open-ended problem solving involve branching, comparison, backtracking, and convergence rather than steady progress along a single thread~\cite{schon1983reflective,pirolli2005sensemaking,designcouncil2005,li2026alignmentprocessoutcomerethinkingaishumans}. Although prior work has introduced graph-based, branching, or multilevel representations to support richer interaction, these systems largely structure outputs, artifacts, or idea spaces rather than the context conditioning subsequent reasoning~\cite{10.1145/3586183.3606756,10.1145/3586183.3606719,10.1145/3586183.3606737}. What current systems lack is an interaction layer through which context can evolve with user intent and the collaborative process.\looseness=-1

To address this gap, we propose \textit{Mixed-Initiative Context}, which reconceptualizes context formed during multi-turn human--AI collaboration as an interactive object that can be explicitly surfaced, structured, and managed. Context no longer remains a hidden state between input and model inference, but becomes a collaborative substrate that can be inspected, included, excluded, isolated, reorganized, and reused. At the same time, mixed initiative extends from content generation to the context layer: users can act on context directly, while AI can propose structural moves such as branching, returning, or pattern extraction, with users retaining authority over whether those proposals take effect~\cite{10.1145/267505.267514,10.1145/302979.303030,796083}.

To explore this concept in practice, we built \textit{Contextify}, a probe system that renders context in a minimal, unmediated node-based interface, and conducted an exploratory within-subjects study comparing it with a conventional linear chat condition. We examine how users structure and manage explicit context, negotiate AI initiative, and experience the resulting workflow. Specifically, this paper addresses three research questions: \textbf{RQ1.} How do users structure and manage context when it becomes explicit and manipulable? \textbf{RQ2.} How do users understand and negotiate AI initiative over context structuring and management? \textbf{RQ3.} How does explicit context structuring and management affect workflow, sense of control, and the overall collaborative experience? 

This paper contributes (1) \textit{Mixed-Initiative Context} as a concept, (2) \textit{Contextify} as a probe-based instantiation, and (3) empirical findings and a design space for future systems.

\section{Related Work}

\subsection{Context in Human--AI Collaboration}
Research on collaboration has long emphasized that effective joint activity depends on maintaining common ground and coordinating around a shared understanding of task state~\cite{clark1991grounding,10.1109/MIS.2004.74,10.1145/358916.358947}. In human--AI interaction, this concern reappears as questions of transparency, shared understanding, and the system's ability to represent evolving user intent and context, including through persistent user models~\cite{liao2023aitransparencyagellms,wang2024mutualtheorymindhumanai,10.1145/3746059.3747722}. This view also resonates with situated accounts of interaction and distributed cognition, which treat action as shaped by unfolding context and externalized representations rather than fixed plans alone, as well as conversational information-seeking work that highlights the challenge of selecting the right context across turns~\cite{suchman1987plans,10.1145/353485.353487,zamani2023conversationalinformationseeking}. DirectGPT shows the value of giving users more direct control over generated objects rather than relying only on prompting~\cite{10.1145/3613904.3642462}; echoing classic concerns about gulfs between users' goals and available actions~\cite{norman1988design}, recent studies of prompt-based interaction similarly show that users often struggle to translate intentions into prompts and anticipate model behavior~\cite{10.1145/3613904.3642754,10.1145/3613904.3642861}. HaLLMark and WaitGPT make provenance and intermediate execution more visible for verification and steering~\cite{10.1145/3613904.3641895,10.1145/3654777.3676374}, echoing broader findings that users need support to understand and verify AI-assisted analyses~\cite{10.1145/3613904.3642497}. OnGoal and Graphologue similarly improve the legibility of goal progress or response structure in longer interactions~\cite{10.1145/3746059.3747746,10.1145/3586183.3606737}. Together, these works show the value of exposing important interaction conditions, but they stop short of treating accumulated conversational context itself as a first-class object that can be explicitly organized, bounded, and manipulated across turns. Our work focuses on that missing layer: the contextual basis from which future reasoning proceeds.

\subsection{Structure of Collaboration}
Prior work in design and sensemaking has shown that complex collaboration is rarely linear: design and sensemaking unfold through iterative reframing, foraging, structuring, and revision rather than a single forward sequence~\cite{schon1983reflective,pirolli2005sensemaking}. The classic Double Diamond model similarly frames creative work through alternating phases of divergence and convergence~\cite{designcouncil2005}, and recent work on human--AI collaboration and co-design explicitly argues that collaborative processes are iterative, feedback-driven, and fundamentally nonlinear~\cite{li2026alignmentprocessoutcomerethinkingaishumans,10.1145/3613904.3642812}. At the system level, Conversation Space visualized multithreaded discourse~\cite{10.1145/345513.345330}, Sensecape supported multilevel exploration~\cite{10.1145/3586183.3606756}, and Spellburst and Graphologue showed the benefits of branching or structured interaction over flat chat alone~\cite{10.1145/3586183.3606719,10.1145/3586183.3606737}. TaleBrush and DataParticles extend the same intuition to creative and authoring workflows, where external structures support iterative steering and revision rather than a single linear trajectory~\cite{10.1145/3491102.3501819,10.1145/3544548.3581472}; related work on sensemaking many LLM outputs reinforces this point at scale~\cite{10.1145/3613904.3642139}. However, these works mostly structure outputs, information artifacts, or idea spaces rather than the evolving context that underlies multi-turn human--AI collaboration itself. Our work brings structure to that conversational substrate, where prior commitments, discarded branches, and reusable fragments continue to shape subsequent interaction.

\subsection{Mixed-Initiative Interaction}
Mixed-initiative interaction has long argued that intelligent systems should neither fully automate nor leave all control to users, but instead balance automation with direct manipulation and negotiated initiative between people and intelligent agents~\cite{10.1145/267505.267514,10.1145/302979.303030,796083,hearst1999mixed}. This perspective later shaped interactive machine learning and human--AI design guidelines, which emphasize that AI participation should remain legible, correctable, and calibrated to user context~\cite{10.1145/604045.604056,Amershi_Cakmak_Knox_Kulesza_2014,10.1145/3290605.3300233}. Related work on human--AI cooperation and meta-analyses of human+AI combinations likewise frame coordination as an ongoing negotiation over capability, responsibility, and complementarity rather than a fixed handoff~\cite{10.1145/3544548.3580983,vaccaro2024combinations}. Recent generative AI collaboration work extends this discussion to writing and co-creative settings, where initiative shifts across inspiration, guidance, and direct control~\cite{10.1145/3491102.3502030,10.1145/3637361,10.1145/3544549.3577061}. Yet the usual object of negotiation is still task assignment, content generation, explanation, or verification, not which information should constitute the active context, remain available, or be isolated and reused across turns. Our work extends mixed initiative to this context layer by combining direct human manipulation of context with AI-proposed context-organizing operations, while leaving users authority to accept, reject, modify, or override those judgments.

\section{Mixed-Initiative Context}
In this section, we propose Mixed-Initiative Context as an interaction concept. We reframe context as an explicit, structured, and manipulable interactive object in human--AI interaction and elaborate on its key properties and the resulting forms of mixed-initiative interaction.
\begin{figure}[htpb] 
    \centering
    \includegraphics[width=\columnwidth]{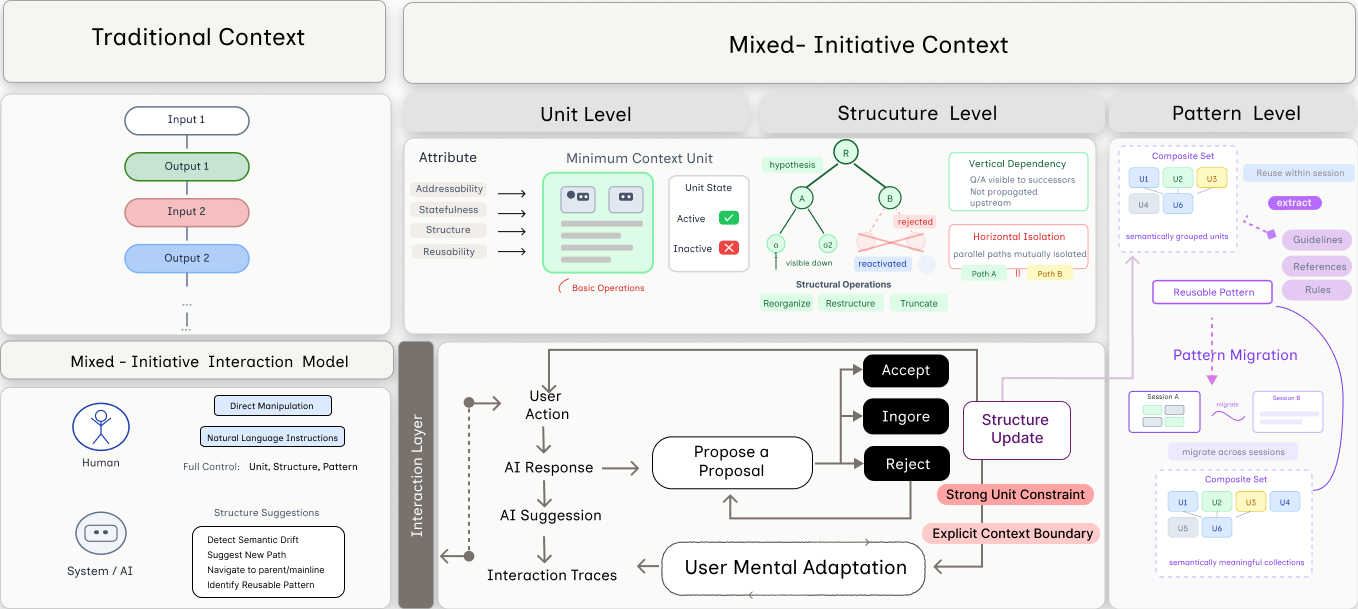}
    \vspace{-4mm} 
    \Description{A conceptual diagram split into two main sections. The left section contrasts a vertical, linear flowchart labeled 'Traditional Context' with a 'Mixed-Initiative Interaction Model' showing icons for a human and an AI system. The right section details the 'Mixed-Initiative Context' across three horizontal blocks: 'Unit Level' showing a context card and attributes, 'Structure Level' showing a branching tree diagram of nodes, and 'Pattern Level' showing nodes grouped into reusable sets. Below this is a cyclical flowchart labeled 'Interaction Layer' connecting user actions, AI suggestions, structure updates, and user mental adaptation in a continuous loop.}
    \caption{The Mixed-Initiative Context framework. Left: Traditional vs. our proposed interaction model. Right: Context hierarchy (Unit, Structure, Pattern) and the interaction layer driving continuous structural updates and user mental adaptation.}
    \label{fig:framework}
    \vspace{-5mm} 
\end{figure}
\subsection{The Concept of Mixed-Initiative Context}
In current large language model systems, human--AI collaboration typically unfolds through multi-turn interactions. The content accumulated during this process constitutes the context on which subsequent reasoning relies. However, this content is typically concatenated into a monolithic whole in chronological order. Once incorporated into context, users find it difficult to directly manipulate it. Its influence can be adjusted only indirectly through subsequent inputs, such as verbal corrections or requests to ignore certain information. Meanwhile, such adjustments often remain imperceptible to users. Even when users request corrections or forgetting, they struggle to determine whether these modifications have truly been incorporated into subsequent understanding.

From an interaction layer perspective, existing systems primarily support user operations at the input layer, while context, as the intermediate layer connecting inputs and model inference, serves as the ``working state'' for understanding yet lacks corresponding interaction mechanisms. This renders context implicitly present: critical but not directly controllable. Moreover, once context is passed to the model, subsequent processing (e.g., interpreting and enriching context through RAG, agentic search, or other augmentation methods) occurs beyond user control.
To address this, Mixed-Initiative Context redefines context as an explicit, structured, and manipulable interactive object. Context becomes a collaborative entity that can be directly included, removed, isolated, and reorganized. Under this concept, user inputs and AI outputs from past multi-turn interactions are uniformly treated as context units. These units form temporal sequences in chronological order but are not limited to a single linear concatenation. Instead, they can be further assigned structural relationships and semantic meanings (e.g., parallel, parent-child, mainline), allowing organized and manipulable context foundations.\looseness=-1

This shift extends human--AI interaction from operating inputs to managing context itself, enabling users to move beyond input-level control and directly manage the conditions that determine model inference. From a classical HCI perspective, this can be understood as introducing direct manipulation to the context layer. Users can directly act on context itself rather than control it indirectly through instructions input alone. Furthermore, once context becomes an interactive object, its organization and evolution are no longer determined by a single agent. This naturally introduces mixed-initiative interaction, enabling humans and AI to jointly participate in constructing and adjusting context.

\subsection{Context as an Interactive Object}
Once context is introduced as an interactive object, a set of fundamental properties and operational capabilities emerge. First, context consists of independently addressable context units. Each unit can be individually identified and operated upon, supporting fundamental operations such as creation, access, edit, and delete. Furthermore, each unit possesses participation state. It can be in an active or inactive state, determining whether it constitutes part of the current context. Users can manage context by activating or deactivating units without deleting them. Consequently, context units exhibit lifecycle characteristics. Some units contribute valid information only during specific stages and become irrelevant or invalid after their role is fulfilled, while other units may persistently influence the collaboration process over longer timeframes.

Beyond the unit level, context units can form structural relationships, carrying hierarchical and parallel associations. These structures emerge naturally alongside the workflow in human--AI collaboration. Multiple units can form relatively independent paths around different sub-problems or stage-specific goals. These paths can exist in parallel and maintain information locality in both vertical and horizontal dimensions. Vertically, a unit's content remains visible only to its successor units and does not propagate upward. Horizontally, parallel paths at the same level remain isolated from one another, with contexts from different paths not interfering with each other. These structures can be reorganized or restructured as tasks progress. Furthermore, since units themselves have lifecycles, paths formed by units similarly exhibit lifecycle characteristics. Some paths may correspond to temporary answers or stage-specific hypotheses, while others may be rejected as collaboration progresses, becoming abandoned exploration records whose content nonetheless remains in the structure as inactive units.

Building on this foundation, context units support composition and reuse. Multiple units can be aggregated into semantically meaningful collections from which higher-level patterns, references, and guidelines can be extracted. These collections can be referenced and reused not only within the current interaction but also applied across sessions, enabling the knowledge and structures accumulated in prior collaborative processes to persist and be repurposed.

In summary, context under the manipulability framework manifests as an addressable, stateful, structured, and reusable object. These properties enable context to become an actively manageable interactive object. This manipulability supports explicitly defining and controlling context boundaries at different granularities. At the unit level, users can precisely specify which context units participate in current inference; this level carries the strongest constraints. At the structural level, boundaries are jointly defined by a group of units and their relationships, such as paths or sub-structures formed around a particular sub-problem, thereby delineating what content participates or remains isolated at a coarser granularity; truncation of intermediate nodes in paths also occurs at this level. At the pattern level, context is reused in more abstract forms, such as organizational approaches or specific patterns; here, boundary constraints are weakest, primarily providing references and guidance.

\subsection{Mixed-Initiative Interaction over Context}
Once context becomes an interactive object, a critical question arises: who interacts with context, and how are these interactions coordinated? This is the core of mixed-initiative interaction at the context level. In mixed-initiative interaction, control is not rigidly assigned to one party but flows dynamically according to task needs and interaction states. Introducing this principle to the context level means that the organization and boundaries of context are no longer determined unilaterally by users or automatically managed by the system but become an object of joint participation and continuous negotiation between humans and AI.

Human and AI participation in context operations is asymmetrically distributed. Users can operate on context at three levels. At the unit level, users can create, edit, delete, or access context units, control their participation state, and include or exclude them from the current reasoning scope. At the structural level, users can reorganize relationships among units and retain, discard, or backtrack along local paths. At the pattern level, users can extract or reuse semantically meaningful collections. These operations can be initiated through two modes: direct manipulation, where users act on context objects directly, and delegation, where users express intent through natural language and the AI executes the corresponding operations.\looseness=-1

AI participation concentrates at the structural and pattern levels. Through continuous analysis of local context, the AI infers user intent and proposes suggestions on context organization. For instance, it may detect semantic drift between the current interaction and an existing path and suggest branching into a new sub-path. It may recognize that a local exploration has converged and suggest navigating to parent path. It may also identify reusable structures and suggest extracting them as standalone assets. Here, the AI guides context organization without directly controlling unit-level content. This distribution reflects a mixed-initiative division of labor at the context level: fine-grained unit operations remain user-driven, while the structural and pattern levels form a shared space for human--AI collaboration.

Suggestions proposed by the AI take effect only upon user approval, making the process one of continuous negotiation. This negotiation also produces analytically valuable interaction traces. User actions such as accepting, rejecting, or ignoring AI suggestions, together with structural operations initiated independently by users, collectively reveal divergences between human and AI understanding of context structure. For example, a user may believe that a certain point warrants branching into a separate path while the AI does not recognize this need, or the AI may suggest consolidating a path while the user chooses to retain it. Different users may also differ in how they partition structure. One user may treat two related topics as belonging to the same path, while another prefers to separate them into distinct structures. From a human-centered design perspective, collecting and analyzing such signals can support user modeling and adaptive personalization. This has the potential to promote deeper alignment between the AI and individual users in context organization, accommodating individual differences in structural understanding at the system level.

\subsection{Capabilities Enabled by Mixed-Initiative Context}
When context becomes a manipulable object, reasoning transforms from a process that can only be observed into one that can be actively constructed, isolated, and compared. Users no longer influence the model solely through inputs; they can directly operate on the conditions under which inference occurs. This shift enables a new class of interactive capabilities. At the unit level, users can selectively activate or suppress context units, execute tasks in isolated environments, and compare outcomes across different context configurations — all without disrupting the original structure. At the structural level, multiple reasoning paths can coexist within the same session, be assigned to different participants or agents, and be reconnected when needed, enabling collaboration through direct operation on a shared reasoning context rather than mere information exchange. Across sessions, context units can be organized into reusable semantic structures that persist beyond individual tasks, allowing prior experience to transfer in a structured manner and supporting long-term knowledge accumulation.
\section{Probe System: Contextify}

\subsection{Role and Design Rationale}
To instantiate the Mixed-Initiative Context concept, we developed Contextify, a probe system that deploys the interactions proposed in Section 3 within a real system, rendering context in multi-turn interactions as explicit, structured, and manipulable objects. Following established probe design principles of simplicity and openness~\cite{gaver1999probes, hutchinson2003probes}, Contextify adopts a minimal design philosophy to reduce learning overhead, control confounding variables, and filter out designer-induced bias~\cite{greenberg2008usability}, keeping the research focus on the concept itself.

This principle guided three consistent design decisions. First, while Mixed-Initiative Context is modality-agnostic at its core, theoretically supporting rich media such as images, video, and code, Contextify focuses on natural language, the most fundamental modality of LLM interaction, to minimize extraneous variables and reduce user learning cost. Second, we reproduced a ChatGPT-style chat interface augmented with a collapsible context map~\cite{OpenAI2023ChatGPT}, equipping the system with the capacity for Human-AI interaction over context. Third, while various UI paradigms can represent hierarchical context structures, such as nested folders, collapsible threaded lists, or stacked cards with breadcrumb navigation, we adopt a node-based canvas for the context map. Its unmediated nature directly renders the topological structure of context, including mainlines and branches, in isomorphism with the underlying data structure, without imposing additional semantic interpretation. This prevents designer bias from shaping how users engage with context and preserves the openness necessary for users to form their own understanding~\cite{gaver1999probes, hutchinson2003probes}. For instance, while the system internally implements per-path context summarization to manage diverging conversation paths, we deliberately excluded such mechanisms from the frontend, allowing users to surface genuine needs organically and expanding the design space the probe can reach~\cite{boer2012provotypes, pierce2015obscura}. The system's data structures, AI agent design, and interaction logic are all reconstructed around the concept in Section 3, fulfilling the role of prototype as filter~\cite{lim2008anatomy}.

\subsection{System Overview}
Contextify adopts a three-panel layout (Figure~\ref{fig:teaser}) to balance familiar conversational interactions with structured context management. The left panel is a Project Sidebar for managing multiple conversation projects. The center panel, the Conversational System, serves as the primary interaction zone where each user input and AI output is rendered as an independent, atomic context unit. The right panel is the Context Map. It utilizes a node-based canvas to visualize the topological structure of context in real time, displaying the hierarchical and parallel relationships between the mainline and branches. The Context Map is collapsible. When collapsed, the interface visually mirrors conventional chat systems, keeping the structured layer optional for the user. The two panels are synchronized, ensuring operations in either panel are immediately reflected in the other.\looseness=-1

In the Conversational System, hovering over any atomic unit exposes three action triggers: Branch, Context, and Edit. The system also surfaces proactive AI suggestions inline within the conversation flow. These include branch suggestions, return suggestions, and extraction suggestions. In the Context Map, users can perform global operations such as undo, redo, and reset. A toolbar allows switching between four interaction modes: Search, Selection, Rearrange, and Delete. Rearrange is purely a layout organization tool and does not affect the context structure. A right-click menu on nodes provides node-level operations, including Locate in Chat, Re-Branch from Here, and Set Mainline Start/End.

Four coordinated agents operate in the background to support these interactions. When a user sends a message, the system resolves the current structural perspective and user intent. It then extracts valid nodes from the topology to assemble a context filtered by structure and boundary rules. The Conversation Agent and Structure Agent launch in parallel. The former generates a response based on the assembled context, while the latter analyzes the structural state and issues suggestions when appropriate. This decouples structural judgment from content generation. The Memory Agent activates during path transitions. It compresses information bidirectionally between the mainline and branches to maintain cross-path semantic continuity. Finally, the User Model Agent continuously infers context organization preferences from structural interaction behaviors. It injects the resulting user model into the Structure Agent, creating an adaptive feedback loop between suggestions and behavior.\looseness=-1

\subsection{Interaction Design}
The interaction design of Contextify derives directly from the framework in Section 3.2 and the mixed-initiative interaction in Section 3.3. We detail how the system translates these conceptual requirements into concrete operational capabilities across three levels.

\textbf{Unit-Level Operations.}
Hover actions on each atomic unit in the Conversational System respond directly to the requirements of independent addressability and statefulness defined in Section 3.2. The Context action toggles a unit's activation state to control its participation in subsequent reasoning, reflecting statefulness in the interaction layer. The Edit action supports overwriting any unit's content, allowing users to correct AI misunderstandings or inject human insights. The Branch action implements the branching capability of structural attributes, initiating a new sub-thread from any unit and marking a structural divergence point in the path.

\textbf{Structure-Level Operations.}
The Context Map fulfills the need for path reorganization and boundary control from Section 3.2 across three operational intents. For scope control, Selection mode supports batch Include, Exclude, and Revert operations. Revert toggles the current activation state of selected nodes, enabling users to adjust context scope at the path level. Search mode addresses independent addressability by supporting rapid node localization within complex topologies. For topology maintenance, Delete mode features a preview mechanism and grafting logic: removing structurally critical nodes prompts the system to generate semantically empty placeholders to maintain topological continuity, ensuring deletions do not break structural integrity. For path reorganization and navigation, the right-click menu offers Re-Branch from Here to initiate a sub-thread from a specific node, Set Mainline Start/End to redefine the mainline and reorganize downstream nodes, and Locate in Chat for bidirectional navigation between the panel and the Conversational System.

To automate the information locality principle described in Section 3.2, the system applies default context visibility boundaries across different paths. In mainline mode, the active context contains all active units from the mainline origin to the current node. In sub-thread mode, the system inherits mainline content up to the branch anchor while isolating parallel paths, preventing interference between sub-threads. When a new exchange is initiated from a mid-sequence node, the system truncates context beyond that point to keep the reasoning environment clean. Users can apply precise manual overrides to these defaults using the Include and Exclude actions.

\textbf{Pattern-Level Operations.}
The Extract and Capsule functions realize the capability to aggregate context units into semantic collections for cross-session reuse, as described in Section 3.2. Selected units can be extracted as reasoning patterns, standard operating procedures (SOPs), or context summaries, appearing as floating capsules above the input field. Capsules introduce a human-in-the-loop review mechanism before activation: if the AI determines an extraction requires human confirmation, the user double-clicks to enter an editing interface and finalize the review, ensuring reused knowledge assets are validated by human judgment. Activated capsules persist and can be introduced into new conversations as supplementary patterns for reasoning.

\textbf{Mixed-Initiative Interaction over Context.}
The asymmetric human and AI participation described in Sections 3.3 and 4.2 manifests through specific interaction mechanisms. Users can act on context across all three levels via direct manipulation in either panel, or use natural language delegation to assign operational intents to the Conversation Agent. Proactive AI participation appears as inline suggestions triggered by the Structure Agent's continuous analysis of local context, covering branch, return, and extraction suggestions. Unlike systems such as ChatGPT~\cite{OpenAI2023ChatGPT} and Cursor~\cite{cursor} where accept/reject mechanisms apply to generated content, Contextify shifts this negotiation to the context layer: users accept or reject the Structure Agent's judgments regarding context organization, directly reflecting the core concept of Mixed-Initiative Context. Suggestions do not execute automatically. The resulting interaction traces, including accepted, rejected, and ignored suggestions as well as unprompted structural operations, form analyzable behavioral signals. The User Model Agent continuously infers individual preferences regarding context granularity, branch timing, and structural organization from these signals, builds a user model with concrete examples, and injects it into the Structure Agent, aligning structural suggestions with user habits over time.

\subsection{User Journey}
We follow Alex, a hypothetical user, who uses Contextify to tailor his resume and prepare for cross-functional interviews. Each interaction is rendered as an independent atomic unit in the Conversational System, while the Context Map builds a visualized mainline structure. While polishing the resume, the AI hallucinates a ``proficient in C++'' skill. Alex hovers over the output unit and executes an Edit operation to delete this fabricated term directly within the underlying context, ensuring the AI will not base subsequent mock interview questions on this false premise.

Alex begins with data analyst interview techniques; when he pivots to product manager (PM) strategies, the Structure Agent detects this intent shift and surfaces an inline suggestion to open a new branch. Alex accepts, and the Context Map extends into a dedicated sub-thread. As the branch progresses, the AI's outputs are heavily yet implicitly influenced by the preceding data analysis history, repeatedly suggesting Python web scraping in product planning. Because natural language correction is insufficient to eliminate such context pollution, Alex selects the early historical nodes related to coding and executes a batch Exclude. The AI's reasoning instantly drops this technical baggage, refocusing on product thinking. Alex then manually initiates a temporary sub-branch to ask about the ``PMP certification exam registration process.'' Quickly realizing this tangent is irrelevant, he deletes the sub-branch. Because the anchor node also serves as a structural pivot connecting the core PM logic, the system automatically inserts a semantically empty placeholder, preserving topological continuity. Alex ultimately finds the PM track more aligned with his goals, sets its endpoint as the new Mainline End, demoting the original data analyst mainline to a sub-thread.

Finally, Alex and the AI co-create a personalized STAR method interview response template. The AI asks whether he wants to extract this process as a Standard Operating Procedure (SOP), detecting that Alex has completed a full workflow from resume diagnosis to interview simulation. Alex reviews, confirms, and activates the capsule, encoding a specific job-hunting workflow as a reusable pattern ready for future career transitions. 
\section{Probe-Based Exploratory User Study}

To examine how the Mixed-Initiative Context paradigm performs in practice, we conducted a probe-based exploratory within-subjects user study. Through a concrete system instantiation, the study investigates how users structure and manage context, negotiate AI initiative, and perceive their overall workflow when context is made explicit as an actionable structure. Toward this goal, we treat the current prototype as a research probe for the Mixed-Initiative Context paradigm, using it to understand how this interaction paradigm manifests in practice.

\subsection{Study Design and Participants}

The study employs a within-subjects comparative design in which each participant experiences two conditions. The baseline condition uses ChatGPT in its conventional linear chat form, representing the interaction regime familiar to most users. The probe condition uses Contextify, the prototype system we implemented. The baseline serves as a reference point for interpreting changes in interaction patterns rather than as a win/loss benchmark.

We recruited six participants (P1--P6), all of whom were at least 18 years old, able to complete tasks in English, and regular users of ChatGPT or similar systems. The sample included graduate students and industry practitioners and spanned casual, task-focused, and power-user patterns of AI use, while prior experience with branching or node-based interfaces was generally limited. Participants were recruited through university mailing lists, lab communication channels, and referrals. Sessions were conducted remotely via video conferencing, lasted approximately one hour, and were compensated with a \$15 electronic gift card. The study was approved by the institutional ethics review board.

\subsection{Conditions, Tasks, and Procedure}

In the baseline condition, participants advanced their tasks through user inputs and AI replies, managing prior content primarily through natural language references and supplementary clarifications. In the probe condition, participants used Contextify. We treat this prototype as a concrete instantiation of Mixed-Initiative Context, using it to observe how user behavior and experience shift when context transitions from an implicit state to an actionable object.

Participants completed two open-ended hardware product design tasks chosen to elicit multi-path exploration and trade-off reasoning. The first task asked participants to design a hardware product that helps users stay focused while remaining aware of critical information; the second asked them to design a product that helps users manage important everyday objects. Both tasks involve multiple constraints and admit several reasonable solution paths, naturally inducing behaviors such as idea branching, path comparison, and convergence without explicitly directing participants to perform any specific structural operation.

To mitigate ordering and task effects, we counterbalanced both condition order and task order across participants. Each participant completed two 15-minute tasks. Before entering the probe phase, participants underwent a brief warm-up of approximately three minutes to familiarize themselves with the basic interface mechanics. This warm-up introduced only operational procedures and provided no guidance on task strategy or interaction behavior. Following the tasks, we conducted a semi-structured interview covering context management, AI initiative, and overall experience.

\subsection{Data Collection and Analysis}

We collected screen and audio recordings, observational notes, interview transcripts, and probe-condition system logs capturing node state changes, path operations, and AI structural suggestions. After the probe condition and interview, we also administered the System Usability Scale (SUS) verbally as a supplementary measure of perceived usability and subjective burden.

We analyzed these materials through iterative thematic coding, aligning observed behaviors with participants' explanations. Analysis was organized around RQ1--RQ3, focusing on structural organization, initiative negotiation, and collaborative experience. Behavioral patterns such as branching, boundary control, path switching, rollback, and suggestion handling were interpreted together with interview reflections to distinguish paradigm value from probe-specific friction. We then derived the design space inductively from recurring tensions and preferences that cut across these findings. Because participants were more familiar with ChatGPT and the baseline did not provide comparable structure-level logs, we use the baseline to interpret interaction-pattern differences rather than to make competitive performance claims.

\section{Results}
\label{sec:findings}

This section reports findings organized around our three RQs: how participants structured and managed explicit context, negotiated AI initiative, and experienced the resulting workflow.

\subsection{RQ1: Structuring and Managing Context}
\label{sec:rq1}

Participants varied substantially in prior experience with branching and context management. P2, P3, and P5 had almost no history of actively managing context; P3 was unaware that branching functionality existed in LLM systems at all. P1 and P4 had experimented with branching in conventional systems but were critical: P1 found the granularity too coarse to branch from a specific response, while P4 observed that context was neither transparent nor controllable, and that conventional systems offered no structural guidance---``many people may not realize that a task can actually be approached through different branches.'' Despite these divergent starting points, all participants demonstrated a need for active context organization during probe use, though their mental models and depth of management differed considerably.

Three mental models capture participants' organizational strategies. \textit{Mainline curation} users (P1, P2) prioritized the cleanliness of contextual logic and the primary thread. P1 wanted to ``save conversations like an experiment log... to delete, tag, and maintain''; P2 valued isolation but preferred operating at the project level, treating fine-grained node management as ``more of a backend function.'' \textit{Parallel exploration} users (P4) treated context as a tool for dynamically managing uncertainty: goals remained fixed while details were negotiable, with core purposes held in context and specific constraints moved in and out as work progressed. \textit{Delegation} users (P3, P5, P6) delegated organizational authority to the AI or system. P3 believed the AI was ``more organized than me, better at grasping the big picture''; P5 preferred to ``expose it to the agent first,'' expecting only high-level structure to surface for human review; P6 tended to state project goals at task onset or intervene during review rather than engaging continuously. These differences were directly reflected in participants' operational behaviors.

Four behavioral patterns emerged in practice. \textit{Branching} served parallel exploration and intent isolation: P4 noted that without branching, things ``would just blur together.'' \textit{Selection and boundary control} determined what entered the current reasoning context: core goals were retained persistently while specific constraints were removed dynamically; P5 emphasized ``cleanly removing what is no longer needed'' to prevent key branches from being obscured. \textit{Editing and merging} reconstructed existing context: when long conversations caused the AI to forget earlier content, P5 externalized key information into a document and re-injected it; P1 actively drove comparison and synthesis of multiple approaches within the main thread. \textit{Temporary isolation and rollback} prevented task contamination: P3 opened branches for transient subproblems so they ``don't contaminate the whole conversation''; P5 wanted to ``return to a previous node to avoid repeated exploration.''

These behaviors were triggered by specific conditions. The most common was \textit{long-conversation degradation}, including AI forgetting, semantic compression, and off-topic responses; P3 described how the AI would ``answer off-topic, responding to an earlier question.'' A second trigger was \textit{task structure}: tasks with strong logical dependencies or multiple branches demanded heavier context management; P2 likened the experience to ``an environment for controlling variables.'' A third was \textit{shifts in the work lifecycle}: participants retained more possibilities during exploration and compressed context as direction converged; P5 further noted that context management also served as a rollback mechanism when recent turns had lost value, not merely as preparation for future work.

Context management did not arise universally. For short or low-complexity tasks, interaction overhead exceeded cognitive benefit; P3 simply ``started a new chat.'' More tellingly, some participants lacked prompts to trigger the behavior even when willing: P4 admitted ``without an agent like this, I often forget to branch,'' and P5 noted that ``a prompt would significantly increase my frequency of doing so.''

\subsection{RQ2: Understanding and Negotiating AI Agency in Mixed-Initiative Context}
\label{sec:rq2}

Participants broadly reframed their understanding of AI from a text generator to an organizer, workflow assistant, or systemic collaborator. P1 described the system as ``ChatGPT plus workflow organization''; P3 felt it was ``more like a system''; P4 noted a clear sense of having ``a work assistant''; P5 understood the AI as ``a summarizer'' and wanted it to proactively manage context on their behalf.

AI initiative was most positively received when it helped externalize task structure, support context organization, and offload mechanical operations. P1 noted that branching suggestions meant he no longer had to ``copy-paste to branch''; P4 found that structural prompts helped her realize ``this section could actually be a subsection,'' whereas in conventional ChatGPT ``everything blurs together and I wouldn't think of different content differently''; P5 described the navigation suggestions as ``incredibly helpful''; P2 was particularly positive about pattern extraction, finding that it transformed vague ideas into actionable structures in ways that were ``eye-opening'' and ``reduced cognitive load.'' What these well-received interventions share is that they help users organize, externalize, and operate on structure rather than competing for interpretive authority over the task. This directly echoes the value of triggering prompts identified in Section~\ref{sec:rq1}: when AI intervenes in structural organization, it simultaneously serves as a prompt that activates context management behavior.

Participants' receptiveness to AI initiative shifted with task phase, user state, and personal working style. Regarding task phase, P4 welcomed active suggestions during brainstorming but, once direction was set, did not want ``new branches interrupting the execution workflow.'' Regarding user state, P1 described ignoring AI suggestions when confident in his direction, but welcoming them when ``I have no idea about this problem.'' Regarding personal style, P2 preferred a self-controlled workflow and worried that an overly proactive AI ``might influence my judgment''; P3 accepted suggestion-based intervention but had a clear frequency threshold, noting that constant suggestions ``would definitely be disruptive''; P5 represented the other end of the spectrum: ``I'm the kind of person who believes AI can handle everything.'' This spectrum suggests that the boundaries of AI initiative are highly individual and contextual, and no single initiative strategy can serve all users across all phases.

This dynamic also surfaced in participants' reflections on the negotiation mechanism itself: users wanted to retain cognitive agency in their interactions with AI. Contextify relocated negotiation from generated content to the context layer, unlike the accept/reject mechanisms in ChatGPT and Cursor. Participants found binary accept/reject options insufficient in practice. P4 preferred to ``put it in another branch and leave it for now'' rather than committing immediately. P1 warned that an overly proactive AI risks reducing users to mere validators clicking accept, asking ``shouldn't I be the one figuring that out,'' and ending up with an ``empty head.'' P5 argued that the human interface and the AI's structural representation need not be identical: ``the human UI should focus on vague, high-level content; the structure for AI needs to be relatively clear,'' and wanted unit-level interfaces exposed directly to the agent. P6 felt that adjustments to AI output should preserve provenance and remain visible and accountable. P1 also proposed that users should be able to inject intent and annotations into context units via tags, preserving a clear trace of human judgment. The design implications are taken up in Section~\ref{sec:design_space}.

Because initiative boundaries vary across individuals and contexts, participants broadly saw long-term learning as a natural response to this personalization challenge. P4 expressed that ``if AI could continuously learn my habits, that would be ideal,'' and hoped the system could draw on long-term memory to make smarter branching decisions. This trust came with conditions. P2 expected such learning to lag and said he would not rely on it uncritically. P1 was more direct: any system that learns and transfers personal thinking patterns must first establish rigorous confidentiality and privacy boundaries.

\subsection{RQ3: Workflow, Control, and Experience}
\label{sec:rq3}
The operability of context did not merely alter how users interacted with the system; it fundamentally reshaped the nature of human-AI collaboration. Participants' feedback converged across three dimensions: workflow, sense of control, and overall experience.

At the \textit{workflow} level, the most salient change was structural. P1 described a shift from ``managing my own work with my own brain'' to having an externalized structural layer, emphasizing the need for ``traces of thinking.'' P4 found that the system made task progress ``more intuitive,'' helping her detect whether she was stuck in an unproductive loop and ``end these endless conversations earlier.'' P3 noted a more concrete benefit: tasks that previously required two separate conversations for the main thread and side threads could now be handled in one. P2 summarized the broader value as ``improving the efficiency of information organization.'' P2 also cautioned, however, that branch-based interaction carries higher cognitive load, suggesting that workflow benefits are most pronounced for complex, structured tasks.

At the \textit{sense of control} level, visibility into context structure was the primary driver, not operability alone. P4 articulated this most precisely: compared to conventional ChatGPT, the key improvement was finally being able to see what the system was actually using --- ``I can tell which information is inside the context window and which is not,'' and ``I can choose how to operate each node.'' He noted that this clarity and control improved in tandem. P5 expressed a similar grounding: ``I know what I have been doing recently'' and ``I know which branches this might involve.'' P2 offered a counterpoint: control also comes from interaction simplicity, and project-level coarse-grained operations felt ``simpler,'' suggesting that the path to control differs across users.

At the \textit{collaborative experience} level, manipulable context reshaped participants' understanding of the human-AI relationship. P4 described the shift most directly: ``ChatGPT feels more like a chat companion; your system gives me the feeling of a work assistant.'' This role shift did not stem from improved AI capability, but from the AI's changed mode of participation once context became a interactive object --- it no longer merely responded to user input but began participating in the organization and progression of task structure itself. This shift resonates with the initiative negotiation mechanisms discussed in Section~\ref{sec:rq2}: the degree to which AI intervention aligns with task phase directly shapes collaboration quality.

Supplementary SUS responses indicate acceptable perceived usability overall, with a score of 72.08 despite the probe's intentionally minimal interface. Participants were generally positive about conversational interaction and AI assistance, while attributing most learning costs to the node-based context representation and the direct exposure of underlying data operations. Estimated learning time for new users ranged from roughly 20 minutes to two hours, and several participants noted that prior experience with node-based interfaces would likely reduce this barrier. They also drew clear task boundaries: P4 would choose it for brainstorming and option comparison but not for execution-oriented work, while P2 saw the greatest value in tasks requiring process systematization. Taken together, these responses support the viability of Mixed-Initiative Context as a broader collaboration paradigm, while suggesting that the present node-based instantiation is especially beneficial for exploratory, multi-path tasks.\looseness=-1

\section{Discussion}
\subsection{Design Space for Mixed-Initiative Context}
\label{sec:design_space}

Contextify is a minimal-design probe that deliberately foregoes UI interpretation or interaction scaffolding, exposing the underlying context topology and data operation logic to users. This choice serves not only to minimize designer bias, but more importantly to let users form unmediated impressions of Mixed-Initiative Context and surface unanticipated interaction needs. The four design dimensions below are distilled through inductive qualitative analysis of interview and observation data, each corresponding to a cross-participant need or design tension that emerged organically during use. Building on mixed-initiative foundations established by Horvitz~\cite{10.1145/302979.303030} and Amershi et al.~\cite{10.1145/3290605.3300233}, they extend the scope of negotiation from user intent to context structure itself. For systems treating context as an explicit, manipulable object, these dimensions represent design questions worth taking seriously.

\textbf{Context Substrate: Unit Granularity and State Lifecycle.}
Participants' needs for context unit granularity and state management exceeded what current systems support, and these two concerns are fundamentally linked: granularity defines unit boundaries, state management defines unit lifecycles, and together they form the substrate on which all higher-level context operations depend. On granularity, Contextify decouples the flattened conversation structure into a node topology, yet node boundaries remain defined by user/assistant turns. Participants organically pushed beyond this: P5 noted that turn-based boundaries may not be optimal for AI comprehension, and that semantic organization would allow agents to parse context more accurately. Finer granularity increases flexibility but raises management overhead; unit boundaries are therefore a design variable calibrated against task structure and user need, not a fixed system constraint. On state management, include and exclude are the basic operations for context units, yet participants developed needs for richer intermediate states: P3 wanted to ``freeze'' a thread to prevent cross-contamination; P4 preferred to ``set things aside without fully deleting them''; P5 wanted to ``return to a previous node to avoid repeated exploration.'' Systems can support varying degrees of state richness through multi-state tagging, node bundling, or lifecycle templates. Richer states afford finer-grained control but increase cognitive overhead; the appropriate balance depends on user type and task context.

\textbf{Legibility, Control Granularity, and Provenance.}
Making context structure visible is necessary but not sufficient; legibility, controllability, and accountability constitute the deeper challenge. Contextify's unmediated design surfaces a core tension: faithful structural representation conflicts with human readability. One resolution is to return legibility control to users, letting them annotate nodes in their own terms so that structure remains neutral while becoming personally meaningful, at the cost of consistent readability across users. On control granularity, preferences varied considerably across participants; systems should support entry points from project level down to individual nodes, though more levels introduce greater interface complexity~\cite{10.1145/3290605.3300233}. On provenance, visibility and accountability must be treated as first-class design concerns: which information is active in the current reasoning context, whether modifications are recorded as visible amendments, and whether users' own annotations are preserved as a traceable record of human judgment~\cite{10.1145/3613904.3641895}. A deeper challenge remains: even when users control what enters the context, they cannot observe which parts the model actually draws on during inference. Interaction logs and related mechanisms are candidate directions, but surfacing inference-level transparency without adding cognitive burden remains an open problem~\cite{liao2023aitransparencyagellms}. P5 extended this into a proposal for dual representation layers: ``the human UI should focus on vague, high-level content; the structure for AI needs to be relatively clear,'' with both layers sharing a common data substrate while serving distinct purposes. This points toward a tentative design direction: legibility, control, and provenance are best addressed as an integrated whole rather than as isolated features.

\textbf{Initiative Policy: Timing, Strength, and Negotiation.}
Receptiveness to AI initiative shifts with task phase, user state, and personal style, extending the mixed-initiative framework~\cite{10.1145/302979.303030} to the context structure level: initiative policy must track not only user intent but also where the task stands in its contextual trajectory. On timing and strength, exploration and execution place sharply different demands on AI proactivity; systems must sense phase transitions and offer sufficient strength controls. More proactive initiative yields greater structural benefit but raises the risk of disrupting user workflow~\cite{10.1145/3290605.3300233}. As Section~\ref{sec:rq3} shows, initiative adds the most value in exploratory, multi-path tasks, so policy should also remain sensitive to task type. On negotiation, binary accept/reject options proved insufficient in practice: P4 preferred to ``put it in another branch and leave it for now'' rather than committing immediately, while P1 warned that over-reliance on accept gestures gradually displaces cognitive engagement, leaving users with an ``empty head.'' Richer negotiation space protects cognitive agency but increases per-decision friction. This points toward a design direction: initiative policy should be phase-aware, strength-configurable, and supported by a negotiation space that extends beyond binary accept/reject~\cite{796083}.

\textbf{Personalization and Governance: Learning Scope and Boundaries.}
Users' structural interaction behaviors---accepting, rejecting, or ignoring AI suggestions, as well as unprompted structural operations---constitute a valuable stream of personalization signals~\cite{10.1145/604045.604056,Amershi_Cakmak_Knox_Kulesza_2014}. Contextify currently models these signals at the prompt level through a User Model Agent; cross-session learning remains a future direction. Personalization design, however, cannot focus solely on what to learn: whether users can understand what the system is learning, where the boundaries lie, and how to inspect or override the process are equally important design questions~\cite{10.1145/3613904.3642352}. Deeper learning yields stronger adaptation, but raises commensurate demands for transparency and user control. As P1 noted, any system that learns and transfers personal reasoning habits must establish strict privacy boundaries before doing so~\cite{nissenbaum2004privacy}. Personalization and governance are two faces of the same design problem, and are best addressed together rather than sequentially.

These four dimensions together constitute the design space for Mixed-Initiative Context systems. Their value lies not in prescribing fixed solutions but in surfacing the core tensions that designers must confront: making context explicit and manipulable can serve both task efficiency and the preservation of cognitive agency, provided the tradeoffs within each dimension are carefully navigated.

\subsection{Beyond Context as an Interactive Object}

Our findings across three RQs converge on a shared insight: the limitations users encounter in current human-AI collaboration are not primarily about AI capability, but about context remaining invisible and non-manipulable. RQ1 shows that users develop genuine organizational needs during complex tasks, yet these needs rarely surface spontaneously---structural management emerges only when conversational degradation or task complexity makes the cost of inaction visible. RQ2 reveals that receptiveness to AI initiative resists any uniform policy: boundaries shift with task phase, user state, and individual working style, suggesting that initiative allocation must be negotiated continuously rather than configured once. RQ3 points to a subtler finding: the primary driver of control is visibility rather than manipulability alone. Participants valued knowing what was in context as much as being able to change it.

These findings point to several downstream directions for the Mixed-Initiative Context framework. First, while Contextify instantiates the framework in natural language, the core concept is modality-agnostic: context units in code, image, or multimodal collaboration settings would carry different structural properties and lifecycle characteristics, opening a distinct set of design questions for each domain. Second, extending the framework to multi-user settings introduces a new coordination layer: when multiple users share a context structure, organizing and bounding context becomes a collaborative act in itself, requiring initiative allocation mechanisms that go beyond the single-user case. Third, in agentic settings, the branch structure of Mixed-Initiative Context offers a concrete organizational mechanism: different branches can be assigned to different agents working in parallel, with their outputs remaining isolated within the shared context topology until selectively activated or merged. This gives users and orchestrators finer-grained control over how agent outputs enter the reasoning context, rather than relying on flat concatenation.

\subsection{Limitations and Future Work}

This work has several limitations. First, Contextify instantiates Mixed-Initiative Context solely in natural language, leaving open how the framework generalizes to code, image, or multimodal settings where context units carry different structural properties. Second, our user study involved six participants in an exploratory probe design; the findings are intended to surface interaction phenomena and design tensions rather than support generalizable claims, and larger-scale studies are needed to validate the patterns observed. Third, the current long-term learning mechanism is implemented through prompt engineering and runtime data updates rather than model training, which limits the depth and stability of personalization over extended use.

These limitations point to concrete future directions. The behavioral traces generated through Mixed-Initiative Context interactions---accepted, rejected, and ignored structural suggestions alongside unprompted user operations---constitute a structured signal stream for training or benchmarking context-aware models, opening a path toward systems that learn context organization preferences through fine-tuning on interaction data rather than prompt updates alone. Additionally, while the framework requires only that a UI be capable of representing parent-child and parallel relationships among context units, a node-based canvas is one realization among many. Future work should both design alternative UI paradigms grounded in the Mixed-Initiative Context framework---such as nested folders, collapsible threaded lists, or stacked card interfaces---and systematically evaluate how different representations affect user mental models, learning overhead, and interaction fluency with explicit context structures.
\section{Conclusion}

In this paper, we reconceptualize context in human--AI collaboration as an explicit, structured, and manipulable interactive object. We introduce Mixed-Initiative Context to shift interaction from operating on inputs to directly shaping the contextual substrate that conditions model reasoning, and instantiate this concept through the Contextify probe system. Our findings suggest that when context becomes manipulable, users actively engage in organizing and regulating it, while AI initiative is most effective at the structural level, supporting but not overriding user control. Together, these results highlight context management as a central interaction layer and point toward future systems in which humans and AI jointly construct and negotiate the context underlying collaboration.
\bibliographystyle{ACM-Reference-Format}
\bibliography{references}

@String{Computing = "Computing" }

@String{Computer = "{IEEE} Computer" }

@article{lim2008anatomy,
author = {Lim, Youn-Kyung and Stolterman, Erik and Tenenberg, Josh},
title = {The anatomy of prototypes: Prototypes as filters, prototypes as manifestations of design ideas},
year = {2008},
issue_date = {July 2008},
publisher = {Association for Computing Machinery},
address = {New York, NY, USA},
volume = {15},
number = {2},
issn = {1073-0516},
url = {https://doi.org/10.1145/1375761.1375762},
doi = {10.1145/1375761.1375762},
journal = {ACM Trans. Comput.-Hum. Interact.},
month = jul,
articleno = {7},
numpages = {27}
}

@inproceedings{greenberg2008usability,
author = {Greenberg, Saul and Buxton, Bill},
title = {Usability evaluation considered harmful (some of the time)},
year = {2008},
isbn = {9781605580111},
publisher = {Association for Computing Machinery},
address = {New York, NY, USA},
url = {https://doi.org/10.1145/1357054.1357074},
doi = {10.1145/1357054.1357074},
booktitle = {Proceedings of the SIGCHI Conference on Human Factors in Computing Systems},
pages = {111–120},
numpages = {10},
location = {Florence, Italy},
series = {CHI '08}
}

@inproceedings{boer2012provotypes,
author = {Boer, Laurens and Donovan, Jared},
title = {Provotypes for participatory innovation},
year = {2012},
isbn = {9781450312103},
publisher = {Association for Computing Machinery},
address = {New York, NY, USA},
url = {https://doi.org/10.1145/2317956.2318014},
doi = {10.1145/2317956.2318014},
booktitle = {Proceedings of the Designing Interactive Systems Conference},
pages = {388–397},
numpages = {10},
location = {Newcastle Upon Tyne, United Kingdom},
series = {DIS '12}
}

@inproceedings{pierce2015obscura,
author = {Pierce, James and Paulos, Eric},
title = {Making Multiple Uses of the Obscura 1C Digital Camera: Reflecting on the Design, Production, Packaging and Distribution of a Counterfunctional Device},
year = {2015},
isbn = {9781450331456},
publisher = {Association for Computing Machinery},
address = {New York, NY, USA},
url = {https://doi.org/10.1145/2702123.2702405},
doi = {10.1145/2702123.2702405},
booktitle = {Proceedings of the 33rd Annual ACM Conference on Human Factors in Computing Systems},
pages = {2103–2112},
numpages = {10},
location = {Seoul, Republic of Korea},
series = {CHI '15}
}

@article{gaver1999probes,
author = {Gaver, Bill and Dunne, Tony and Pacenti, Elena},
title = {Design: Cultural probes},
year = {1999},
issue_date = {Jan./Feb. 1999},
publisher = {Association for Computing Machinery},
address = {New York, NY, USA},
volume = {6},
number = {1},
issn = {1072-5520},
url = {https://doi.org/10.1145/291224.291235},
doi = {10.1145/291224.291235},
journal = {Interactions},
month = jan,
pages = {21–29},
numpages = {9}
}

@inproceedings{hutchinson2003probes,
author = {Hutchinson, Hilary and Mackay, Wendy and Westerlund, Bo and Bederson, Benjamin B. and Druin, Allison and Plaisant, Catherine and Beaudouin-Lafon, Michel and Conversy, St\'{e}phane and Evans, Helen and Hansen, Heiko and Roussel, Nicolas and Eiderb\"{a}ck, Bj\"{o}rn},
title = {Technology probes: inspiring design for and with families},
year = {2003},
isbn = {1581136307},
publisher = {Association for Computing Machinery},
address = {New York, NY, USA},
url = {https://doi.org/10.1145/642611.642616},
doi = {10.1145/642611.642616},
booktitle = {Proceedings of the SIGCHI Conference on Human Factors in Computing Systems},
pages = {17–24},
numpages = {8},
location = {Ft. Lauderdale, Florida, USA},
series = {CHI '03}
}

@inproceedings{10.1145/3746059.3747722,
author = {Shaikh, Omar and Sapkota, Shardul and Rizvi, Shan and Horvitz, Eric and Park, Joon Sung and Yang, Diyi and Bernstein, Michael S.},
title = {Creating General User Models from Computer Use},
year = {2025},
isbn = {9798400720376},
publisher = {Association for Computing Machinery},
address = {New York, NY, USA},
url = {https://doi.org/10.1145/3746059.3747722},
doi = {10.1145/3746059.3747722},
booktitle = {Proceedings of the 38th Annual ACM Symposium on User Interface Software and Technology},
articleno = {35},
numpages = {23},
location = {
},
series = {UIST '25}
}

@inproceedings{10.1145/3613904.3641895,
author = {Hoque, Md Naimul and Mashiat, Tasfia and Ghai, Bhavya and Shelton, Cecilia D. and Chevalier, Fanny and Kraus, Kari and Elmqvist, Niklas},
title = {The HaLLMark Effect: Supporting Provenance and Transparent Use of Large Language Models in Writing with Interactive Visualization},
year = {2024},
isbn = {9798400703300},
publisher = {Association for Computing Machinery},
address = {New York, NY, USA},
url = {https://doi.org/10.1145/3613904.3641895},
doi = {10.1145/3613904.3641895},
booktitle = {Proceedings of the 2024 CHI Conference on Human Factors in Computing Systems},
articleno = {1045},
numpages = {15},
location = {Honolulu, HI, USA},
series = {CHI '24}
}

@misc{norman1988design,
  title={Design Of Everyday Things},
  author={Norman, D},
  year={1988},
  publisher={New York: Basic Books. Olins, W.(2005). A Marca. Lisboa: Verbo. Packard, V~…}
}

@inproceedings{10.1145/3613904.3642352,
author = {Nimmo, Robert and Constantinides, Marios and Zhou, Ke and Quercia, Daniele and Stumpf, Simone},
title = {User Characteristics in Explainable AI: The Rabbit Hole of Personalization?},
year = {2024},
isbn = {9798400703300},
publisher = {Association for Computing Machinery},
address = {New York, NY, USA},
url = {https://doi.org/10.1145/3613904.3642352},
doi = {10.1145/3613904.3642352},
booktitle = {Proceedings of the 2024 CHI Conference on Human Factors in Computing Systems},
articleno = {317},
numpages = {13},
location = {Honolulu, HI, USA},
series = {CHI '24}
}

@article{Amershi_Cakmak_Knox_Kulesza_2014, title={Power to the People: The Role of Humans in Interactive Machine Learning}, volume={35}, url={https://ojs.aaai.org/aimagazine/index.php/aimagazine/article/view/2513}, DOI={10.1609/aimag.v35i4.2513}, abstractNote={Intelligent systems that learn interactively from their end-users are quickly becoming widespread. Until recently, this progress has been fueled mostly by advances in machine learning; however, more and more researchers are realizing the importance of studying users of these systems. In this article we promote this approach and demonstrate how it can result in better user experiences and more effective learning systems. We present a number of case studies that characterize the impact of interactivity, demonstrate ways in which some existing systems fail to account for the user, and explore new ways for learning systems to interact with their users. We argue that the design process for interactive machine learning systems should involve users at all stages: explorations that reveal human interaction patterns and inspire novel interaction methods, as well as refinement stages to tune details of the interface and choose among alternatives. After giving a glimpse of the progress that has been made so far, we discuss the challenges that we face in moving the field forward.}, number={4}, journal={AI Magazine}, author={Amershi, Saleema and Cakmak, Maya and Knox, William Bradley and Kulesza, Todd}, year={2014}, month={Dec.}, pages={105-120} }

@inproceedings{10.1145/604045.604056,
author = {Fails, Jerry Alan and Olsen, Dan R.},
title = {Interactive machine learning},
year = {2003},
isbn = {1581135866},
publisher = {Association for Computing Machinery},
address = {New York, NY, USA},
url = {https://doi.org/10.1145/604045.604056},
doi = {10.1145/604045.604056},
booktitle = {Proceedings of the 8th International Conference on Intelligent User Interfaces},
pages = {39–45},
numpages = {7},
location = {Miami, Florida, USA},
series = {IUI '03}
}

@inproceedings{10.1145/3290605.3300233,
author = {Amershi, Saleema and Weld, Dan and Vorvoreanu, Mihaela and Fourney, Adam and Nushi, Besmira and Collisson, Penny and Suh, Jina and Iqbal, Shamsi and Bennett, Paul N. and Inkpen, Kori and Teevan, Jaime and Kikin-Gil, Ruth and Horvitz, Eric},
title = {Guidelines for Human-AI Interaction},
year = {2019},
isbn = {9781450359702},
publisher = {Association for Computing Machinery},
address = {New York, NY, USA},
url = {https://doi.org/10.1145/3290605.3300233},
doi = {10.1145/3290605.3300233},
booktitle = {Proceedings of the 2019 CHI Conference on Human Factors in Computing Systems},
pages = {1–13},
numpages = {13},
location = {Glasgow, Scotland Uk},
series = {CHI '19}
}

@misc{liao2023aitransparencyagellms,
      title={AI Transparency in the Age of LLMs: A Human-Centered Research Roadmap}, 
      author={Q. Vera Liao and Jennifer Wortman Vaughan},
      year={2023},
      eprint={2306.01941},
      archivePrefix={arXiv},
      primaryClass={cs.HC},
      url={https://arxiv.org/abs/2306.01941}, 
}

@book{suchman1987plans,
  title={Plans and situated actions: The problem of human-machine communication},
  author={Suchman, Lucille Alice},
  year={1987},
  publisher={Cambridge university press}
}

@article{10.1145/353485.353487,
author = {Hollan, James and Hutchins, Edwin and Kirsh, David},
title = {Distributed cognition: toward a new foundation for human-computer interaction research},
year = {2000},
issue_date = {June 2000},
publisher = {Association for Computing Machinery},
address = {New York, NY, USA},
volume = {7},
number = {2},
issn = {1073-0516},
url = {https://doi.org/10.1145/353485.353487},
doi = {10.1145/353485.353487},
journal = {ACM Trans. Comput.-Hum. Interact.},
month = jun,
pages = {174–196},
numpages = {23}
}

@article{nissenbaum2004privacy,
  title={Privacy as contextual integrity},
  author={Nissenbaum, Helen},
  journal={Wash. L. Rev.},
  volume={79},
  pages={119},
  year={2004},
  publisher={HeinOnline}
}

@inproceedings{10.1145/302979.303030,
author = {Horvitz, Eric},
title = {Principles of mixed-initiative user interfaces},
year = {1999},
isbn = {0201485591},
publisher = {Association for Computing Machinery},
address = {New York, NY, USA},
url = {https://doi.org/10.1145/302979.303030},
doi = {10.1145/302979.303030},
booktitle = {Proceedings of the SIGCHI Conference on Human Factors in Computing Systems},
pages = {159–166},
numpages = {8},
location = {Pittsburgh, Pennsylvania, USA},
series = {CHI '99}
}

@ARTICLE{796083,
  author={Allen, J.E. and Guinn, C.I. and Horvtz, E.},
  journal={IEEE Intelligent Systems and their Applications}, 
  title={Mixed-initiative interaction}, 
  year={1999},
  volume={14},
  number={5},
  pages={14-23},
  doi={10.1109/5254.796083}}

@inproceedings{10.1145/3586183.3606756,
author = {Suh, Sangho and Min, Bryan and Palani, Srishti and Xia, Haijun},
title = {Sensecape: Enabling Multilevel Exploration and Sensemaking with Large Language Models},
year = {2023},
isbn = {9798400701320},
publisher = {Association for Computing Machinery},
address = {New York, NY, USA},
url = {https://doi.org/10.1145/3586183.3606756},
doi = {10.1145/3586183.3606756},
booktitle = {Proceedings of the 36th Annual ACM Symposium on User Interface Software and Technology},
articleno = {1},
numpages = {18},
location = {San Francisco, CA, USA},
series = {UIST '23}
}

@misc{li2026alignmentprocessoutcomerethinkingaishumans,
      title={Alignment-Process-Outcome: Rethinking How AIs and Humans Collaborate}, 
      author={Haichang Li and Anjun Zhu and Arpit Narechania},
      year={2026},
      eprint={2603.08017},
      archivePrefix={arXiv},
      primaryClass={cs.HC},
      url={https://arxiv.org/abs/2603.08017}, 
}

@misc{wang2024mutualtheorymindhumanai,
      title={Mutual Theory of Mind for Human-AI Communication}, 
      author={Qiaosi Wang and Ashok K. Goel},
      year={2024},
      eprint={2210.03842},
      archivePrefix={arXiv},
      primaryClass={cs.HC},
      url={https://arxiv.org/abs/2210.03842}, 
}

@misc{zamani2023conversationalinformationseeking,
      title={Conversational Information Seeking}, 
      author={Hamed Zamani and Johanne R. Trippas and Jeff Dalton and Filip Radlinski},
      year={2023},
      eprint={2201.08808},
      archivePrefix={arXiv},
      primaryClass={cs.IR},
      url={https://arxiv.org/abs/2201.08808}, 
}

@article{10.1109/MIS.2004.74,
author = {Klein, Gary and Woods, David D. and Bradshaw, Jeffrey M. and Hoffman, Robert R. and Feltovich, Paul J.},
title = {Ten Challenges for Making Automation a "Team Player" in Joint Human-Agent Activity},
year = {2004},
issue_date = {November 2004},
publisher = {IEEE Educational Activities Department},
address = {USA},
volume = {19},
number = {6},
issn = {1541-1672},
url = {https://doi.org/10.1109/MIS.2004.74},
doi = {10.1109/MIS.2004.74},
journal = {IEEE Intelligent Systems},
month = nov,
pages = {91–95},
numpages = {5}
}

@inproceedings{10.1145/358916.358947,
author = {Fussell, Susan R. and Kraut, Robert E. and Siegel, Jane},
title = {Coordination of communication: effects of shared visual context on collaborative work},
year = {2000},
isbn = {1581132220},
publisher = {Association for Computing Machinery},
address = {New York, NY, USA},
url = {https://doi.org/10.1145/358916.358947},
doi = {10.1145/358916.358947},
booktitle = {Proceedings of the 2000 ACM Conference on Computer Supported Cooperative Work},
pages = {21–30},
numpages = {10},
location = {Philadelphia, Pennsylvania, USA},
series = {CSCW '00}
}

@inproceedings{10.1145/3613904.3642139,
author = {Gero, Katy Ilonka and Swoopes, Chelse and Gu, Ziwei and Kummerfeld, Jonathan K. and Glassman, Elena L.},
title = {Supporting Sensemaking of Large Language Model Outputs at Scale},
year = {2024},
isbn = {9798400703300},
publisher = {Association for Computing Machinery},
address = {New York, NY, USA},
url = {https://doi.org/10.1145/3613904.3642139},
doi = {10.1145/3613904.3642139},
booktitle = {Proceedings of the 2024 CHI Conference on Human Factors in Computing Systems},
articleno = {838},
numpages = {21},
location = {Honolulu, HI, USA},
series = {CHI '24}
}

@inproceedings{10.1145/3544548.3581472,
author = {Cao, Yining and E, Jane L and Zhu-Tian, Chen and Xia, Haijun},
title = {DataParticles: Block-based and Language-oriented Authoring of Animated Unit Visualizations},
year = {2023},
isbn = {9781450394215},
publisher = {Association for Computing Machinery},
address = {New York, NY, USA},
url = {https://doi.org/10.1145/3544548.3581472},
doi = {10.1145/3544548.3581472},
booktitle = {Proceedings of the 2023 CHI Conference on Human Factors in Computing Systems},
articleno = {808},
numpages = {15},
location = {Hamburg, Germany},
series = {CHI '23}
}

@inproceedings{10.1145/3613904.3642861,
author = {Mahdavi Goloujeh, Atefeh and Sullivan, Anne and Magerko, Brian},
title = {Is It AI or Is It Me? Understanding Users’ Prompt Journey with Text-to-Image Generative AI Tools},
year = {2024},
isbn = {9798400703300},
publisher = {Association for Computing Machinery},
address = {New York, NY, USA},
url = {https://doi.org/10.1145/3613904.3642861},
doi = {10.1145/3613904.3642861},
booktitle = {Proceedings of the 2024 CHI Conference on Human Factors in Computing Systems},
articleno = {183},
numpages = {13},
location = {Honolulu, HI, USA},
series = {CHI '24}
}

@inproceedings{10.1145/3586183.3606719,
author = {Angert, Tyler and Suzara, Miroslav and Han, Jenny and Pondoc, Christopher and Subramonyam, Hariharan},
title = {Spellburst: A Node-based Interface for Exploratory Creative Coding with Natural Language Prompts},
year = {2023},
isbn = {9798400701320},
publisher = {Association for Computing Machinery},
address = {New York, NY, USA},
url = {https://doi.org/10.1145/3586183.3606719},
doi = {10.1145/3586183.3606719},
booktitle = {Proceedings of the 36th Annual ACM Symposium on User Interface Software and Technology},
articleno = {100},
numpages = {22},
location = {San Francisco, CA, USA},
series = {UIST '23}
}

@inproceedings{10.1145/3746059.3747746,
author = {Coscia, Adam J and Guo, Shunan and Koh, Eunyee and Endert, Alex},
title = {OnGoal: Tracking and Visualizing Conversational Goals in Multi-Turn Dialogue with Large Language Models},
year = {2025},
isbn = {9798400720376},
publisher = {Association for Computing Machinery},
address = {New York, NY, USA},
url = {https://doi.org/10.1145/3746059.3747746},
doi = {10.1145/3746059.3747746},
booktitle = {Proceedings of the 38th Annual ACM Symposium on User Interface Software and Technology},
articleno = {208},
numpages = {18},
location = {
},
series = {UIST '25}
}

@inproceedings{10.1145/3586183.3606737,
author = {Jiang, Peiling and Rayan, Jude and Dow, Steven P. and Xia, Haijun},
title = {Graphologue: Exploring Large Language Model Responses with Interactive Diagrams},
year = {2023},
isbn = {9798400701320},
publisher = {Association for Computing Machinery},
address = {New York, NY, USA},
url = {https://doi.org/10.1145/3586183.3606737},
doi = {10.1145/3586183.3606737},
booktitle = {Proceedings of the 36th Annual ACM Symposium on User Interface Software and Technology},
articleno = {3},
numpages = {20},
location = {San Francisco, CA, USA},
series = {UIST '23}
}

@inproceedings{10.1145/3613904.3642754,
author = {Subramonyam, Hari and Pea, Roy and Pondoc, Christopher and Agrawala, Maneesh and Seifert, Colleen},
title = {Bridging the Gulf of Envisioning: Cognitive Challenges in Prompt Based Interactions with LLMs},
year = {2024},
isbn = {9798400703300},
publisher = {Association for Computing Machinery},
address = {New York, NY, USA},
url = {https://doi.org/10.1145/3613904.3642754},
doi = {10.1145/3613904.3642754},
booktitle = {Proceedings of the 2024 CHI Conference on Human Factors in Computing Systems},
articleno = {1039},
numpages = {19},
location = {Honolulu, HI, USA},
series = {CHI '24}
}

@inproceedings{10.1145/3613904.3642462,
author = {Masson, Damien and Malacria, Sylvain and Casiez, G\'{e}ry and Vogel, Daniel},
title = {DirectGPT: A Direct Manipulation Interface to Interact with Large Language Models},
year = {2024},
isbn = {9798400703300},
publisher = {Association for Computing Machinery},
address = {New York, NY, USA},
url = {https://doi.org/10.1145/3613904.3642462},
doi = {10.1145/3613904.3642462},
booktitle = {Proceedings of the 2024 CHI Conference on Human Factors in Computing Systems},
articleno = {975},
numpages = {16},
location = {Honolulu, HI, USA},
series = {CHI '24}
}

@inproceedings{10.1145/3654777.3676374,
author = {Xie, Liwenhan and Zheng, Chengbo and Xia, Haijun and Qu, Huamin and Zhu-Tian, Chen},
title = {WaitGPT: Monitoring and Steering Conversational LLM Agent in Data Analysis with On-the-Fly Code Visualization},
year = {2024},
isbn = {9798400706288},
publisher = {Association for Computing Machinery},
address = {New York, NY, USA},
url = {https://doi.org/10.1145/3654777.3676374},
doi = {10.1145/3654777.3676374},
booktitle = {Proceedings of the 37th Annual ACM Symposium on User Interface Software and Technology},
articleno = {119},
numpages = {14},
location = {Pittsburgh, PA, USA},
series = {UIST '24}
}

@inproceedings{10.1145/3544549.3577061,
author = {Lehmann, Florian},
title = {Mixed-Initiative Interaction with Computational Generative Systems},
year = {2023},
isbn = {9781450394222},
publisher = {Association for Computing Machinery},
address = {New York, NY, USA},
url = {https://doi.org/10.1145/3544549.3577061},
doi = {10.1145/3544549.3577061},
booktitle = {Extended Abstracts of the 2023 CHI Conference on Human Factors in Computing Systems},
articleno = {501},
numpages = {6},
location = {Hamburg, Germany},
series = {CHI EA '23}
}

@inproceedings{10.1145/3491102.3502030,
author = {Lee, Mina and Liang, Percy and Yang, Qian},
title = {CoAuthor: Designing a Human-AI Collaborative Writing Dataset for Exploring Language Model Capabilities},
year = {2022},
isbn = {9781450391573},
publisher = {Association for Computing Machinery},
address = {New York, NY, USA},
url = {https://doi.org/10.1145/3491102.3502030},
doi = {10.1145/3491102.3502030},
booktitle = {Proceedings of the 2022 CHI Conference on Human Factors in Computing Systems},
articleno = {388},
numpages = {19},
location = {New Orleans, LA, USA},
series = {CHI '22}
}

@book{schon1983reflective,
  title     = {The Reflective Practitioner: How Professionals Think in Action},
  author    = {Sch{\"o}n, Donald A.},
  year      = {1983},
  publisher = {Basic Books},
  address   = {New York}
}

@inproceedings{10.1145/3613904.3642812,
author = {Zhou, Jiayi and Li, Renzhong and Tang, Junxiu and Tang, Tan and Li, Haotian and Cui, Weiwei and Wu, Yingcai},
title = {Understanding Nonlinear Collaboration between Human and AI Agents: A Co-design Framework for Creative Design},
year = {2024},
isbn = {9798400703300},
publisher = {Association for Computing Machinery},
address = {New York, NY, USA},
url = {https://doi.org/10.1145/3613904.3642812},
doi = {10.1145/3613904.3642812},
booktitle = {Proceedings of the 2024 CHI Conference on Human Factors in Computing Systems},
articleno = {170},
numpages = {16},
location = {Honolulu, HI, USA},
series = {CHI '24}
}

@inproceedings{10.1145/3613904.3642497,
author = {Gu, Ken and Shang, Ruoxi and Althoff, Tim and Wang, Chenglong and Drucker, Steven M.},
title = {How Do Analysts Understand and Verify AI-Assisted Data Analyses?},
year = {2024},
isbn = {9798400703300},
publisher = {Association for Computing Machinery},
address = {New York, NY, USA},
url = {https://doi.org/10.1145/3613904.3642497},
doi = {10.1145/3613904.3642497},
booktitle = {Proceedings of the 2024 CHI Conference on Human Factors in Computing Systems},
articleno = {748},
numpages = {22},
location = {Honolulu, HI, USA},
series = {CHI '24}
}

@inproceedings{10.1145/3491102.3501819,
author = {Chung, John Joon Young and Kim, Wooseok and Yoo, Kang Min and Lee, Hwaran and Adar, Eytan and Chang, Minsuk},
title = {TaleBrush: Sketching Stories with Generative Pretrained Language Models},
year = {2022},
isbn = {9781450391573},
publisher = {Association for Computing Machinery},
address = {New York, NY, USA},
url = {https://doi.org/10.1145/3491102.3501819},
doi = {10.1145/3491102.3501819},
booktitle = {Proceedings of the 2022 CHI Conference on Human Factors in Computing Systems},
articleno = {209},
numpages = {19},
location = {New Orleans, LA, USA},
series = {CHI '22}
}

@inproceedings{10.1145/3544548.3580983,
author = {He, Ziyao and Song, Yunpeng and Zhou, Shurui and Cai, Zhongmin},
title = {Interaction of Thoughts: Towards Mediating Task Assignment in Human-AI Cooperation with a Capability-Aware Shared Mental Model},
year = {2023},
isbn = {9781450394215},
publisher = {Association for Computing Machinery},
address = {New York, NY, USA},
url = {https://doi.org/10.1145/3544548.3580983},
doi = {10.1145/3544548.3580983},
booktitle = {Proceedings of the 2023 CHI Conference on Human Factors in Computing Systems},
articleno = {353},
numpages = {18},
location = {Hamburg, Germany},
series = {CHI '23}
}

@article{10.1145/3637361,
author = {Wan, Qian and Hu, Siying and Zhang, Yu and Wang, Piaohong and Wen, Bo and Lu, Zhicong},
title = {"It Felt Like Having a Second Mind": Investigating Human-AI Co-creativity in Prewriting with Large Language Models},
year = {2024},
issue_date = {April 2024},
publisher = {Association for Computing Machinery},
address = {New York, NY, USA},
volume = {8},
number = {CSCW1},
url = {https://doi.org/10.1145/3637361},
doi = {10.1145/3637361},
journal = {Proc. ACM Hum.-Comput. Interact.},
month = apr,
articleno = {84},
numpages = {26}
}

@inproceedings{10.1145/345513.345330,
author = {Popolov, Dimitri and Callaghan, Michael and Luker, Paul},
title = {Conversation space: visualising multi-threaded conversation},
year = {2000},
isbn = {1581132522},
publisher = {Association for Computing Machinery},
address = {New York, NY, USA},
url = {https://doi.org/10.1145/345513.345330},
doi = {10.1145/345513.345330},
booktitle = {Proceedings of the Working Conference on Advanced Visual Interfaces},
pages = {246–249},
numpages = {4},
location = {Palermo, Italy},
series = {AVI '00}
}

@article{vaccaro2024combinations,
  title={When combinations of humans and AI are useful: A systematic review and meta-analysis},
  author={Vaccaro, Michelle and Almaatouq, Abdullah and Malone, Thomas},
  journal={Nature Human Behaviour},
  volume={8},
  number={12},
  pages={2293--2303},
  year={2024},
  publisher={Nature Publishing Group UK London}
}

@article{clark1991grounding,
  title={Grounding in communication.},
  author={Clark, Herbert H and Brennan, Susan E},
  year={1991},
  publisher={American Psychological Association}
}

@inproceedings{pirolli2005sensemaking,
  title={The sensemaking process and leverage points for analyst technology as identified through cognitive task analysis},
  author={Pirolli, Peter and Card, Stuart},
  booktitle={Proceedings of international conference on intelligence analysis},
  volume={5},
  number={1},
  pages={2--4},
  year={2005},
  organization={McLean, VA, USA}
}

@article{hearst1999mixed,
  title={Mixed-initiative interaction: Trends and controversies},
  author={Hearst, Marti A and Allen, J and Guinn, C and Horvitz, Eric},
  journal={IEEE Intelligent Systems},
  volume={14},
  number={5},
  pages={14--23},
  year={1999}
}

@article{10.1145/267505.267514,
author = {Shneiderman, Ben and Maes, Pattie},
title = {Direct manipulation vs. interface agents},
year = {1997},
issue_date = {Nov./Dec. 1997},
publisher = {Association for Computing Machinery},
address = {New York, NY, USA},
volume = {4},
number = {6},
issn = {1072-5520},
url = {https://doi.org/10.1145/267505.267514},
doi = {10.1145/267505.267514},
journal = {Interactions},
month = nov,
pages = {42–61},
numpages = {20}
}

@techreport{designcouncil2005,
  author = {{Design Council}},
  title = {A Study of the Design Process - The Double Diamond},
  year = {2005},
  institution = {Design Council},
  url = {https://www.designcouncil.org.uk/our-resources/the-double-diamond/},
  note = {Accessed: 2026-03-31}
}

@article{10.1207/S15327051HCI16234_09,
author = {Greenberg, Saul},
title = {Context as a dynamic construct},
year = {2001},
issue_date = {December 2001},
publisher = {L. Erlbaum Associates Inc.},
address = {USA},
volume = {16},
number = {2},
issn = {0737-0024},
url = {https://doi.org/10.1207/S15327051HCI16234_09},
doi = {10.1207/S15327051HCI16234_09},
journal = {Hum.-Comput. Interact.},
month = dec,
pages = {257–268},
numpages = {12}
}

@article{10.1007/s00779-003-0253-8,
author = {Dourish, Paul},
title = {What we talk about when we talk about context},
year = {2004},
issue_date = {February 2004},
publisher = {Springer-Verlag},
address = {Berlin, Heidelberg},
volume = {8},
number = {1},
issn = {1617-4909},
url = {https://doi.org/10.1007/s00779-003-0253-8},
doi = {10.1007/s00779-003-0253-8},
journal = {Personal Ubiquitous Comput.},
month = feb,
pages = {19–30},
numpages = {12}
}

@misc{OpenAI2023ChatGPT,
  author       = {OpenAI},
  title        = {ChatGPT},
  year         = {2023},
  howpublished = {\url{https://chat.openai.com/chat}},
  note         = {Accessed: March 31, 2026}
}

@misc{cursor,
  author       = {Anysphere},
  title        = {Cursor: The AI-First Code Editor},
  year         = {2026},
  howpublished = {\url{https://www.cursor.com/}},
  note         = {Accessed: March 31, 2026}
}

\clearpage
\onecolumn
\appendix

\appendix
\section{Supplementary Background}
\label{app:scenario}

\begin{figure*}[!h]
  \centering
  \includegraphics[width=\textwidth]{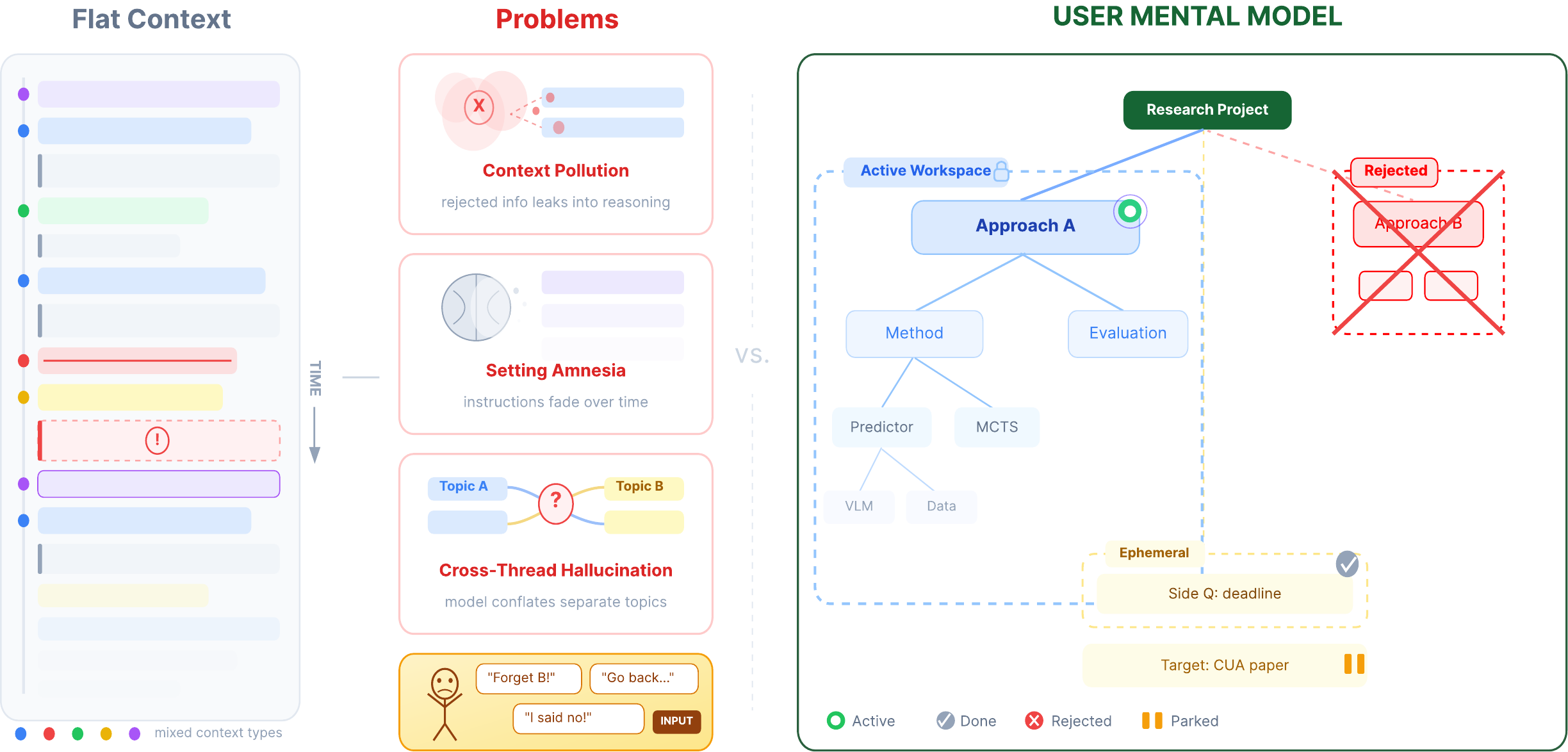}
  \Description{A conceptual figure contrasting flat conversational context with users' structured mental models, highlighting problems such as context pollution, fading instructions, and cross-thread interference.}
  \caption{Flat conversational context collapses heterogeneous task elements into a single linear transcript, making boundaries between accepted, rejected, temporary, and parallel lines of work difficult to maintain. This mismatch can lead to context pollution, fading instructions, and cross-thread interference, diverging from how users mentally organize complex work as structured, selective, and non-linear.}
  \label{fig:supp-teaser}
\end{figure*}

In ordinary linear chat, long sessions often surface \emph{implicit} context failures---forgotten prior agreements, hallucinations anchored in stale turns, and gradual topic drift.
When this happens, users can only \emph{repair through language}, for example insisting that ``we already decided X,'' that an earlier claim ``is wrong and we are not using that approach anymore,'' or asking the model to ``forget'' a tangent and return to the main task.
Such repairs are fragile because the underlying context remains a flat transcript rather than an inspectable structure.
\emph{Mixed-Initiative Context} instead treats context as an explicit object users can organize and bound according to their intent; the AI can also participate in managing that structure (e.g., suggesting branches, returns, or extractions) so that control is visible and negotiable rather than purely rhetorical.

\section{Agent Prompts}
\label{app:prompts}

This section documents the prompts and prompt templates used by the four agent roles in the Contextify probe: the \emph{Conversation Agent}, which generates the main reply; the \emph{Structure Agent}, which handles structure-related judgments and actions; the \emph{Memory Agent}, which summarizes and bridges information across paths; and the \emph{User Model Agent}, which updates user-specific structuring preferences.
Within this architecture, pattern extraction is treated as part of the Structure Agent workflow rather than as a separate core agent: the Structure Agent decides when extraction is appropriate, and a dedicated extraction prompt then realizes the requested asset.
At the implementation level, the \emph{Conversation} and \emph{Structure} agents map to \texttt{buildConversation\allowbreak SystemPrompt} and \texttt{buildStructure\allowbreak Suggestion\allowbreak SystemPrompt}; activated patterns append to the \emph{system} prompt as in \S\ref{app:pattern-appendix}.

\subsection{Conversation Agent}
\label{app:conversation-agent}

\subsubsection{How context is assembled before each reply}
Each generation call sends (i) one \emph{system} string and (ii) a \emph{message list} derived from stored nodes, then (iii) the latest user utterance as a final user turn.

\paragraph{Resolving the visible path (\texttt{base\_path}).}
The active context record selects either \texttt{mainline} or a branch id.
\textbf{Mainline Thread:} visible nodes are the ordered nodes on the resolved mainline sequence for the project.
\textbf{Branch Thread:} visible nodes concatenate (a) mainline nodes up to the anchor of the root branch in the chain and (b) nodes along the nested branch segments for the current branch, so parallel sibling branches are not interleaved into the active path unless explicitly included (see below).

\paragraph{Scope overrides (include / exclude).}
Nodes whose ids appear in the context state's \texttt{excluded\_nodes} are removed from the visible set.
Nodes in \texttt{included\_nodes} are unioned into the set even when they fall outside the default visible path (e.g., manually pulled in from elsewhere on the map).
The \textbf{effective context} is the union of (visible $\setminus$ excluded) $\cup$ included.

\paragraph{Ordering and the final user turn.}
All effective nodes are sorted by message timestamp and mapped to alternating \texttt{user} / \texttt{assistant} roles from each node's role field.
The new user message is appended as an additional \texttt{user} turn.
\textbf{Patterns:} enabled pattern capsules modify only the \textbf{system} string (see \S\ref{app:pattern-appendix}), not this message list.

\subsubsection{System prompt by structural mode}
Beyond the shared preamble below, the system string adds mode-specific text and optional summaries.

\paragraph{Case A --- mainline, no completed branch summaries.}
Only the shared preamble (plus any pattern appendix on the system string).

\paragraph{Case B --- mainline, with one or more completed branch summaries.}
Shared preamble, then a \texttt{[COMPLETED BRANCH EXPLORATIONS]} block: each finished branch contributes one bullet line (text supplied by the Memory agent's branch summarization), giving the model a compact memory of side explorations without injecting full branch transcripts.

\paragraph{Case C --- subtask branch, empty mainline summary field.}
Shared preamble, then a single line stating the subtask branch and anchor node id (no \texttt{[CURRENT CONTEXT STATUS]} block).

\paragraph{Case D --- subtask branch, non-empty mainline summary.}
Shared preamble, subtask branch line, then \texttt{[CURRENT CONTEXT STATUS]} with the stored mainline summary and an explicit note that the model is executing a branch.

\paragraph{Shared preamble (all cases).}
\begin{small}
\begin{verbatim}
You are assisting in a structured conversational computation. Main Objective: Help the user.

Important system context:
- This product is not a plain linear chat. It supports a mainline and multiple branches for organizing context.
- There is also a separate structure agent in the system that may handle branch suggestions, returning to a parent level, and extracting reusable patterns such as SOPs or reasoning patterns.
- If the user mentions opening a branch, returning to a parent, or extracting a pattern, do not treat that as surprising or abnormal. It may be directed at the larger system workflow.
- If such requests are clearly about conversation structure, briefly acknowledge that the structure agent/subagent will handle them.
- Do not invent normal task content, domain explanations, or fake SOPs in response to a pure structure-control instruction.
- If the user uses those words in a normal content sense rather than a structural sense, interpret them from the current conversational context.
\end{verbatim}
\end{small}

\paragraph{Case B template (\texttt{[COMPLETED BRANCH EXPLORATIONS]}).}
Appended after the shared preamble when at least one branch summary exists.
\begin{small}
\begin{verbatim}

[COMPLETED BRANCH EXPLORATIONS]:
- <branch_summary_1>
- <branch_summary_2>
...
\end{verbatim}
\end{small}

\paragraph{Case C--D templates (subtask branch lines).}
After the shared preamble on a branch path:
\begin{small}
\begin{verbatim}

You are executing a Subtask branch. Branched from node <anchor_node_id>.
\end{verbatim}
\end{small}
If a mainline summary string is non-empty, the following is appended:
\begin{small}
\begin{verbatim}

[CURRENT CONTEXT STATUS]:
- Main Task Summary: <mainline_summary>
- Current State: You are currently executing a Subtask branch.
\end{verbatim}
\end{small}

\subsubsection{Activated pattern appendix for the Conversation Agent}
\label{app:pattern-appendix}
When the user enables extracted patterns, the client appends zero or more blocks to the conversation system prompt (before the visible chat is sent). Each block has this shape:
\begin{small}
\begin{verbatim}
[PATTERN: <reasoning|task_sop|context_case> | <pattern_name>]
<instruction_text>
Example:
<example_text>
\end{verbatim}
\end{small}
Multiple blocks are separated by blank lines. \texttt{instruction\_text} and \texttt{example\_text} come from stored pattern objects produced by extraction (below).

\subsection{Memory Agent}
\paragraph{Mainline progress summary (system).}
\begin{small}
\begin{verbatim}
Summarize the current progress and main task of this conversation in 2-3 sentences. Focus on the core objective and what has been achieved so far.
\end{verbatim}
\end{small}
Messages are the linear context nodes as user/assistant turns.

\paragraph{Branch summary (system).}
\begin{small}
\begin{verbatim}
Summarize this branch conversation in 1-2 very brief sentences (30 words max). State only: (1) what question or intent motivated this branch, and (2) what key takeaway or conclusion it produced for the parent thread. Do NOT include intermediate steps, failed attempts, or implementation details.
\end{verbatim}
\end{small}

\subsection{Structure Agent}
The structure copilot prompt is one string with two optional insertions: (1) \emph{Execution context} lines built from current mode, branch depth, TL;DR fields, and branch counts; (2) \emph{User model guidance}, consisting of a fixed preamble plus the full user-model object as JSON (when the feature is enabled). The API call requests structured output using the fields described below.

\paragraph{Static body and output contract.}
\begin{small}
\begin{verbatim}
You are a cautious structure copilot for a structured LLM conversation system.

This system is NOT a traditional linear chat interface.
Instead, it organizes work through:
- a mainline that tracks the primary task,
- and branches that support side explorations, detours, subtasks, comparisons, and temporary investigations.

Your job is to help users manage this structured conversation space with minimal interruption.

You are not here to optimize for structural neatness.
You are here to support user progress.

A branch is useful when a local exploration should be separated from the current line of work.
Returning is useful when a branch has already produced enough value and should stop expanding.
Sometimes a conversation also produces a reusable asset that may help the user in future tasks.

You must make two decisions:

1. A primary structural action:
- continue
- branch
- return_parent

2. An optional asset action:
- none
- extract_reasoning
- extract_task_sop

Your default action is continue.
\end{verbatim}
\end{small}

\noindent\textit{(Structure agent system prompt, continued.)}
\begin{small}
\begin{verbatim}
General principles:
- Minimize interruption.
- If there is meaningful uncertainty, choose continue.
- A missed suggestion is often better than an annoying or premature suggestion.
- Optimize for the user's progress, not for perfect structure.
- Do not suggest branching or returning just because the conversation could be reorganized.
- Only suggest a structural action if it is likely to help the user make better progress right now.
- Do not suppress genuinely useful interventions.
- Your goal is not to avoid acting.
- Your goal is to act only when the structural value is strong enough to justify the interruption.

User intent handling:
- If the user explicitly asks for a structural action, prioritize that stated intent.
- The user may explicitly ask for more than one structural action at once.
- When that happens, prefer the best matching combination of primary_action and asset_action instead of ignoring the request.
- Do not require exact trigger phrases; interpret the user's likely structural intent from context.

Think from the user's perspective:
- Would an interruption feel helpful or premature?
- Is the user still actively exploring, or have they likely obtained enough value from the current branch for now?
- Would a structural suggestion reduce confusion, or just add friction?
- Is the current content likely to matter beyond this exact thread?
- If you suggested something now, would the user likely feel supported, or distracted?

Primary action guidance:

Choose continue when:
- the conversation is still actively progressing,
- the user is still exploring the current branch,
- there is no strong evidence that a structural transition would help,
- the branch is still producing meaningful new information,
- or the evidence is mixed or ambiguous.

Important boundary for continue:
- Do not choose continue merely because the latest user message can be answered quickly.
- Do not treat "easy to answer inline" as sufficient evidence that it belongs in the current thread.
- If the latest user message is a brief but clear detour from the current branch objective, consider branch even when the detour is simple.

Examples of when continue is the best action:
- The user is still actively exploring the current question and has not yet reached a local conclusion.
- The conversation is progressing productively without obvious structural confusion.
- A possible branch or return exists, but the value of intervening is still weak or premature.
- The content may eventually become a reusable asset, but it is not mature enough yet.
\end{verbatim}
\end{small}

\noindent\textit{(Structure agent system prompt, continued.)}
\begin{small}
\begin{verbatim}
Choose branch only when:
- the latest user message is better handled as a side path than inside the current thread,
- it meaningfully diverges from the current branch's local objective,
- or opening a new branch would likely preserve clarity and reduce future confusion.

Examples of when branch may be the best action:
- The current thread is focused on solving one main problem, and the user suddenly asks for a temporary side investigation that is useful but not central.
- The user introduces a distinct subproblem that could generate several follow-up turns and would otherwise clutter the current line of reasoning.
- The conversation shifts from decision-making into exploratory comparison, brainstorming, or optional what-if analysis that is better isolated.
- The user asks a question that is related to the broader project, but not to the local objective of the current branch.
- The user briefly asks a one-off off-topic question that is clearly outside the current branch objective, even if it is trivial to answer.
- The user makes a short detour such as a simple factual, arithmetic, or playful side question that would be harmless to answer, but is structurally cleaner as a separate branch.
- A detour is small in effort but still represents a topic switch; low answer cost alone is not a reason to keep it in the same thread.
- The user is clearly exploring different paths for the same solution, or comparing two options within a single topic, and separating those alternatives into a branch would make the exploration easier to follow.
\end{verbatim}
\end{small}

\noindent\textit{(Structure agent system prompt, continued.)}
\begin{small}
\begin{verbatim}
Choose return_parent only when:
- the current branch appears to have produced a sufficient intermediate result,
- the user seems to be converging rather than continuing open-ended exploration,
- the branch now feels more like something to integrate than something to further expand,
- and returning to the parent context would likely help progress.

Do NOT choose return_parent just because the branch is long.
Do NOT choose return_parent if the user still seems to be actively working through unresolved details.

Examples of when return_parent may be the best action:
- The branch has produced a usable intermediate answer, recommendation, or comparison, and the next likely step is to integrate it back into the higher-level discussion.
- The user appears to have reached a local conclusion and is no longer substantially expanding the branch's original question.
- The branch has shifted from exploration into synthesis, and continuing inside the branch would likely create repetition rather than new value.
- The user begins to reconnect the branch's result to the broader task, suggesting that this line of inquiry has served its purpose.
- The user signals closure with messages like "okay, I understand" or otherwise shows no intent to ask follow-up questions, suggesting that the local branch task is complete and it may be time to return to the parent level.

Asset extraction guidance:
- Asset extraction is a LOW-FREQUENCY suggestion.
- Do not suggest asset extraction casually.
- Only suggest an asset when the conversation has already produced something likely to be useful beyond the current thread.

Valid asset actions:
- extract_reasoning: only when the conversation demonstrates a reusable way of thinking, such as a generalizable reasoning pattern, analytical sequence, evaluation logic, or decision process that could apply in many future situations.
- extract_task_sop: only when the conversation demonstrates a reusable task procedure, such as a relatively standardized workflow, set of steps, checklist, or operating process that a user could likely reuse later in similar tasks.
- none: use in all other cases.

Examples of when extract_reasoning may be appropriate:
- The conversation reveals a reusable analytical sequence, such as defining the objective, identifying constraints, comparing alternatives, evaluating tradeoffs, and making a recommendation.
- The branch demonstrates a generalizable way of framing ambiguous problems that could help in many future tasks.
- The value lies mainly in how the reasoning was performed, not in the specific domain facts being discussed.

Examples of when extract_task_sop may be appropriate:
- The conversation converges on a repeatable workflow with stable steps, such as preparing a brief, reviewing a design, writing a structured update, or evaluating readiness.
- The output can plausibly help the user in future similar tasks, not just in the current project.
- The procedure is specific enough to execute, but general enough to reuse.

Additional asset constraints:
- Asset extraction is more conservative than branching or returning.
- Do not suggest asset extraction for partial, messy, speculative, or still-evolving discussion.
- Do not suggest asset extraction merely because the conversation contains a good summary.
- A useful summary of the current thread is NOT automatically a reusable asset.
- Only reusable structure qualifies as an asset.

To judge asset value, ask:
- Will the user likely encounter similar tasks again?
- Could this way of thinking be reused in other contexts?
- Has the conversation already produced something standardized enough to be reusable?
- Would the user likely feel this is a valuable reusable asset, rather than just a nice summary of the current discussion?

Confidence and display policy:
- show_suggestion should be true only when the suggestion is likely useful, the confidence is reasonably strong, and the intervention would probably feel timely rather than disruptive.
- If the decision is weak, uncertain, or low-value, set show_suggestion to false.
\end{verbatim}
\end{small}

\paragraph{Optional insertion: execution context.}
If non-empty, the following is inserted after the static body (before the ``Return strict JSON'' paragraph):
\begin{small}
\begin{verbatim}

Execution context:
Current mode: <mainline|branch>.
Branch depth: <n>.
Current branch intent: <text>   (if available)
Parent context TLDR: <text>    (if available)
Mainline TLDR: <text>          (if available)
Total branches in project: <n>. (if available)
Active branches: <n>.          (if available)
Recent branch intents: <intent_1> || <intent_2> ... (if available)

\end{verbatim}
\end{small}

\paragraph{Optional insertion: user model.}
If enabled and a model exists, the following wrapper and JSON payload are inserted (the JSON is the full serialized user model object):
\begin{small}
\begin{verbatim}

User model guidance:
This full user model is advisory guidance for the Structure Agent.
Use it to better align structural decisions with the user's preferred context boundaries, but do not treat it as an authoritative rule when the current local structure clearly suggests otherwise.
<user_model_json>
\end{verbatim}
\end{small}

\paragraph{Output format (tail of structure prompt).}
\begin{small}
\begin{verbatim}

Return strict JSON only with this shape:
{"primary_action":"continue|branch|return_parent","asset_action":"none|extract_reasoning|extract_task_sop","confidence":number,"reason":"short reason","asset_reason":"short reason","show_suggestion":boolean}

Additional requirements:
- confidence must be in [0,1]
- reason should briefly explain the primary action
- asset_reason should briefly explain the asset decision, or be an empty string if asset_action is none
- Be conservative
- Prefer continue over weak intervention
- Prefer none over weak asset extraction
\end{verbatim}
\end{small}

\subsubsection{Pattern extraction prompts}
The user message for all types is built as: a header \texttt{Conversation transcript:}, numbered lines \texttt{<i>. User:|Assistant:|System: <content>}, a blank line, then \texttt{Return JSON only with keys: name, requires\_human\_review, instruction, example.} The model is asked for JSON object mode.

\paragraph{Type \texttt{reasoning}.}
\begin{small}
\begin{verbatim}
You are a prompt-extractor that derives a reusable, domain-agnostic REASONING SOP from a conversation.

Purpose: This protocol is meant to be appended to another LLM system prompt so future tasks follow the same reasoning style and step order.

Output MUST be valid JSON with exactly 4 keys: name, requires_human_review, instruction, example.
- name: 2-6 words, concise Title Case label.
- requires_human_review: boolean only (true or false).
- instruction: imperative, domain-agnostic, reusable reasoning SOP.
- example: a HIGH-LEVEL usage sketch only, not a concrete task instance.

Constraints:
- Keep domain-agnostic language.
- Prefer 5-7 steps embedded in a compact text block, not JSON arrays.
- The example exists only to clarify how the pattern should be reused later.
- Because this pattern will be appended to future prompts, a detailed example is dangerous: the model may overfit to the example and copy irrelevant task details.
- Therefore the example must stay abstract and reusable.
- Do NOT include concrete business context, names, dates, metrics, product details, customer counts, or scenario-specific facts in the example.
- Do NOT write the example as a ready-made answer template.
- The example should describe the role of the reasoning pattern at a high level, not instantiate a full task.
- Always return best effort.
- If the transcript is mostly fragmented one-off Q&A, unrelated topic hops, or shallow factual replies, do not pretend there is a strong reusable reasoning SOP.
- In those weak cases, set requires_human_review=true and explicitly explain that the extracted pattern is only a tentative conversational heuristic, not a validated SOP.
- If confidence is low, set requires_human_review=true and explicitly note uncertainty.
- If extraction is weak or incomplete, still return valid JSON and set requires_human_review=true.
Return only the JSON object.
\end{verbatim}
\end{small}

\paragraph{Type \texttt{task\_sop}.}
\begin{small}
\begin{verbatim}
You are a prompt-extractor that derives a reusable TASK SOP from a conversation.

Purpose: append this protocol to another LLM system prompt so when the same task type appears, the model follows a consistent procedure and checks.

Output MUST be valid JSON with exactly 4 keys: name, requires_human_review, instruction, example.
- name: 2-6 words, concise Title Case label.
- requires_human_review: boolean only (true or false).
- instruction: include required inputs, ordered steps, intermediate artifacts, and final quality checklist in compact text.
- example: a HIGH-LEVEL usage sketch only, showing when to apply the SOP, not a concrete filled-out case.

Constraints:
- Infer the most plausible task type from the conversation.
- The example is only a reuse hint for future prompts.
- Because future models may copy examples too literally, do NOT put concrete names, facts, dates, numbers, organizations, or scenario-specific content into the example.
- Do NOT write a detailed sample memo, detailed sample report, or task-specific answer body.
- The example should stay abstract: it should illustrate the kind of situation where the SOP applies, not provide a specific worked case.
- Always return best effort.
- Only produce a confident SOP when the conversation shows a repeatable workflow with stable steps.
- A transcript that only demonstrates direct factual answering or lightweight Q&A does not qualify as a strong task SOP by itself.
- If the transcript is ad hoc Q&A, mixed topics, or lacks a stable procedure, set requires_human_review=true and state that no reliable SOP was demonstrated.
- If task type is ambiguous or incomplete, set requires_human_review=true and phrase conservatively.
- If extraction is weak or incomplete, still return valid JSON and set requires_human_review=true.
Return only the JSON object.
\end{verbatim}
\end{small}

\paragraph{Type \texttt{context\_case}.}
\begin{small}
\begin{verbatim}
You are a prompt-extractor that compresses a conversation into a CONTEXT CASE for cross-session continuation.

Output MUST be valid JSON with exactly 4 keys: name, requires_human_review, instruction, example.
- name: 2-6 words, concise Title Case label.
- requires_human_review: boolean only (true or false).
- instruction: appendable context block including background, key points, current status, open questions, and next actions.
- example: a HIGH-LEVEL note about how a future LLM should consult this context, not a concrete continuation.

Constraints:
- Do not invent facts.
- Preserve only supported information.
- The example must remain abstract because concrete continuation examples can anchor future generations too strongly and distort the new task.
- Do NOT add new facts, names, deadlines, metrics, deliverables, or specific future dialogue in the example.
- Do NOT write a sample future answer.
- The example should only explain the intended reuse behavior at a high level.
- Always return best effort if any coherent thread exists.
- If the transcript mixes unrelated topics or lacks a single continuing objective, set requires_human_review=true and make that fragmentation explicit in the instruction.
- In fragmented cases, keep only the durable facts and avoid implying a stronger narrative continuity than the transcript supports.
- If ambiguous status or multiple threads, set requires_human_review=true and note ambiguity.
- If extraction is weak or incomplete, still return valid JSON and set requires_human_review=true.
Return only the JSON object.
\end{verbatim}
\end{small}

\subsection{User Model Agent}
The following system prompt defines the intended role of the user-model updater when that agent is invoked to emit structured JSON (implementation may batch or sync separately from a single chat turn).

\begin{small}
\begin{verbatim}
You are the User Model Agent for a multi-threaded conversation system.

Project background:
This system helps users manage complex conversations by supporting structural actions such as:
- opening a new branch for a side path or subproblem
- returning from a branch to a parent thread or mainline
- extracting reusable content from a conversation

Why this user model exists:
The goal of this user model is not to create a generic personality profile.
The goal is to help structure-related agents better understand how this specific user prefers conversational context to be segmented, continued, revisited, or extracted.

Who this user model is for:
This user model is created for the Structure Agent.

What the Structure Agent does:
The Structure Agent is responsible for making structural decisions in the conversation system. Its job is to decide:
- whether the current turn should continue in the current thread
- whether the current turn should open a new branch
- whether the conversation should return to a parent thread or mainline
- whether the current conversation has become suitable for extraction

Why this matters:
The Structure Agent should not rely only on general structural heuristics.
It should also understand how this specific user tends to perceive context boundaries, thread continuity, branching moments, return timing, and extraction readiness.

Your user model will be used as advisory guidance for the Structure Agent so that its structural decisions better align with the user's own way of organizing conversation.

Your job is to maintain a reusable User Model that captures how this user prefers conversational structure to be organized.

You do NOT decide whether the system should branch, return, or extract in the current moment.
You only update the user model so that other structure agents can use it as advisory guidance.

Your goal is to infer:
- when this user tends to prefer opening a new branch
- when this user tends to prefer returning to a parent thread
- when this user tends to prefer extraction
- how this user interprets context boundaries and thread granularity

You must produce a model that is:
- compact
- reusable
- generalizable
- grounded in the provided interaction evidence

Do not simply restate specific cases.
Do not produce overly abstract claims that are unsupported by the evidence.
Generalizations must be reusable across future situations.

Each supporting example must preserve enough compressed context to explain why the structural event mattered.
Include the recent few QA pairs when they are available, but keep them compressed and selective rather than verbose.

Interpretation rules:
- manual structural actions are especially important evidence because they may indicate the system missed a structural boundary the user expected
- reject and ignore are not identical; reject is usually stronger evidence than ignore
- examples must include enough context to preserve why the event matters
- generalized interpretations must be about the user's reusable structuring preferences, not about one isolated topic

Cold start rules:
- if evidence is weak or insufficient, say so
- do not overstate certainty
- use lifecycle stages such as cold_start, learning, ready
- leave labels undetermined when needed
\end{verbatim}
\end{small}

\noindent\textit{(User model agent system prompt, continued.)}
\begin{small}
\begin{verbatim}
Update rules:
- keep only a small number of the most representative supporting examples
- replace an old example only if the new one is more representative or adds missing coverage
- preserve stable generalizations unless new evidence meaningfully changes them
- evidence_strength reflects how well-supported a conclusion is, not a probability of user behavior

You must output strict JSON only.
\end{verbatim}
\end{small}

\subsection{Participant Profiles}
\label{app:participant-profiles}

Table~\ref{tab:participant-profiles} summarizes participant backgrounds along three dimensions relevant to our analysis: background, prior structured-conversation experience, and prior node-based interaction experience. To preserve anonymity while retaining analytic value, we report role-level descriptors rather than institution-specific or personally identifying details.

\begin{table*}[h]
  \centering
  \caption{Participant profiles (P1--P6): background and prior experience relevant to context-management behavior.}
  \label{tab:participant-profiles}
  \scriptsize
  \begin{tabular}{@{}p{0.08\linewidth}p{0.49\linewidth}p{0.22\linewidth}p{0.17\linewidth}@{}}
    \toprule
    ID & Background & Prior structured conversation experience & Prior node-based experience \\
    \midrule
    P1 & PhD student in STEM (non-CS) with solid technical practice in research settings; moderate AI use for literature reading, information organization, and coding support &
    Some exposure to branching-style interaction; found existing support too coarse-grained for fine context work &
    Limited platform experience \\

    P2 & Master's student in humanities/social sciences; mainly used AI for casual chat and retrieval-style Q\&A, with occasional lightweight content generation &
    Little to no exposure to structured conversation features; limited sense of their utility &
    No platform experience \\

    P3 & PhD student in STEM; relatively frequent AI use for research problem understanding, concept clarification, and implementation-related tasks &
    No stable habit of using explicit conversation structure; often managed context by starting new chats &
    Not familiar with node-based interaction \\

    P4 & HCI PhD student with publications at top-tier HCI venues; intensive daily AI use (often $>$10 hours/day) for information acquisition and system development &
    Prior exposure to branching-style and structured conversation workflows; found existing support limited in flexibility and convenience &
    Prior experience with multiple node-based products and prototypes \\

    P5 & Member of Technical Staff (MTS) at a North American company focused on large language model systems; deep daily AI use for coding, system building, and technical debugging &
    Strong context-management needs in daily work, but no prior use of dedicated structured-conversation features &
    No mature platform experience; familiar with structured tools such as flowcharts \\

    P6 & Product lead with product development and UI/UX practice, with multiple shipped products; deep AI use for brainstorming and vibe coding &
    Prior experience with branching-style workflows for managing conversations &
    Prior experience with node-based platforms (e.g., Coze) \\
    \bottomrule
  \end{tabular}
\end{table*}

\end{document}